\documentclass[prb,aps,twocolumn,groupedaddress,superscriptaddress,eqsecnum]{revtex4-2}
\usepackage{ulem}
\usepackage[usenames,dvipsnames]{color}
\usepackage[colorlinks=true,linkcolor=Blue,citecolor=Blue,urlcolor=Blue]{hyperref}
\usepackage{graphicx}
\usepackage{verbatim}
\usepackage{bbm}
\usepackage{bm}
\usepackage{amsmath} 
\usepackage{amssymb}
\usepackage{mathptmx}
\usepackage[version=3]{mhchem}
\usepackage{float}

\usepackage{xcolor}
\usepackage{dsfont}

\newcommand{\ba}{\begin{eqnarray}}
\newcommand{\ea}{\end{eqnarray}}
\newcommand{\bd}{\begin{displaymath}}

\newcommand{\nn}{\nonumber \\}

\newcommand{\K}{{\cal K}}

\newcommand{\W}{{\bf W}}
\newcommand{\Z}{{\mathbb Z}} 

\begin{document}
\title{Neural and Tensor-Network Representations of \\ $\mathbb{Z}_N$ Cluster and Dipolar-cluster SPT States}
\author{Seungho \surname{Lee}}
\email{seungho6626@gmail.com} 
\affiliation{Department of Physics, Sungkyunkwan University, Suwon 16419, South Korea}
\affiliation{Institute of Basic Science, Sungkyunkwan University, Suwon 16419, South Korea}
\author{Daesik \surname{Kim}}
\email{kim.ds4811@gmail.com}
\affiliation{Department of Physics, Sungkyunkwan University, Suwon 16419, South Korea}
\author{Hyun-Yong \surname{Lee}}
\email{hyunyong@korea.ac.kr}
\affiliation{Division of Semiconductor Physics, Korea University, Sejong 30019, Korea}
\affiliation{Department of Applied Physics, Graduate School, Korea University, Sejong 30019, Korea}
\author{Jung Hoon \surname{Han}}
\email{hanjemme@gmail.com}
\affiliation{Department of Physics, Sungkyunkwan University, Suwon 16419, South Korea}
\affiliation{Center for Quantum Dynamics of Angular Momentum, Pohang University of Science and Technology, Pohang 37673, South Korea}

\begin{abstract} 
The $\mathbb Z_N$ cluster state is a paradigmatic example of symmetry-protected topological (SPT) order with $\mathbb Z_N\times\mathbb Z_N$ symmetry. We express its wavefunction and those of dipolar cluster states as a projector-based realization of correlator product states, which we dub the $P$-representation. The local projector $P$, also known as the copy tensor, can have multiple legs connecting a site to neighboring sites depending on the model and its deformations. We construct explicit wavefunctions of cluster and dipolar cluster states under deformation using the $P$-representation. The transition to the neural quantum state representations of the same state occurs by splitting the interaction matrices between $\mathbb{Z}_N$ spins into a pair of weight matrices. We derive these weight matrices explicitly for the cluster and dipolar cluster states, thus generalizing previous works on $\mathbb{Z}_2$ cluster states. The projector scheme then leads to the tensor product state (TPS) representation of the cluster-state wavefunctions. We benchmark the resulting TPS construction against conventional MPS representation using density-matrix renormalization group simulations and argue that the TPS could offer efficient representations for some modulated and deformed SPT states.
\end{abstract}
\date{\today}
\maketitle

\section{Introduction}
Representing many-body quantum states efficiently has been a central theme of condensed matter theory for nearly a century. Starting from the Hartree–Fock approximation, successive developments such as Jastrow-correlated wavefunctions, quantum and variational Monte Carlo, exact diagonalization, and density-matrix renormalization group (DMRG) have all contributed essential insights into strongly correlated states of fermions and bosons. With the advent of modern AI and transformer-based methods, the demand for compact yet accurate descriptions of correlated quantum states has only intensified, together with new opportunities to construct such descriptions~\cite{troyer17,becca23}. In particular, representing quantum many-body wavefunctions in the spirit of neural networks—most notably via restricted Boltzmann machines (RBM)—has led to the notion of neural quantum states (NQS)~\cite{troyer17,sarma17a,sarma17b,duan17,cirac18,clark18,xiang18,jia19,jia20,carleo20,carrasquilla20a,aarrasquilla20b,vivas22,bohrdt24,cui25,paul26} which has proven capable of exactly encoding paradigmatic states such as the one-dimensional cluster state and the two-dimensional toric code~\cite{sarma17a,sarma17b,duan17,xiang18,cui25}. In this work, we revisit and systematically generalize the NQS formulation of one-dimensional $\Z_N$ cluster states, focusing on symmetry-protected topological (SPT) order~\cite{wen11,cirac11,pollmann12} and its multipolar extensions~\cite{han24,lam24,han25,you25,aksoy25,ebisu25,bulmash25,pace26,bi26,lam26}, as well as their deformations. In the process, we find it convenient to employ a local projector scheme~\cite{Changlani09,clark18}, referred to as the $P$-representation in this work, as a bridge toward constructing NQS. 

Much progress has been made since the first NQS representation of the $\Z_2$ cluster state appeared~\cite{sarma17a,sarma17b,duan17,xiang18}. The $\Z_2$ cluster state, originally introduced as a universal resource for measurement-based quantum computation~\cite{raussendorf01}, has been generalized to $\Z_N$ cluster states with arbitrary integer $N>2$ and to its variants protected by modulated symmetries—such as dipolar, quadrupolar, and exponential symmetries—which define a broader class of modulated SPT (mSPT) states~\cite{han24,lam24,ebisu25,maeda25,yao25,bulmash25,bi26,lam26} along with the identification of non-invertible Kramers-Wannier symmetries possessed by the various cluster models~\cite{seiberg24,shao24,yamazaki24,han25}. Motivated by these developments, we return to the cluster-state NQS framework with two main goals. First, we extend the NQS construction to uniform $\Z_N$ and dipole-symmetric $\Z_N$ cluster models. This is made possible by providing the first closed-form RBM weights for general $\Z_N$ cluster states. Second, we sharpen the understanding of symmetry fractionalization and other symmetry aspects of these SPT wavefunctions. Concretely, we construct explicit NQS representations for the $\Z_N$ cluster state and for two types of dipole-conserving $\Z_N$ cluster states (type-I and type-II), derive closed-form interaction weight matrices for $\Z_N$ variables, and show how different ways of splitting the weight matrices yield either an MPS or, in the dipolar case, a TPS with three virtual indices per site. In accomplishing these goals, we introduce the $P$-representation and make extensive use of it as a unifying bridge between NQS and tensor-network-type representations.
Using the $P$-representation also helps streamline proofs of push-through conditions, symmetry fractionalization, and writing down matrix-product-operator (MPO) representation of the $\Z_N$ Kramers--Wannier (KW) operator.

Finally, we demonstrate both analytically and numerically by performing DMRG calculations that certain dipolar SPT states admit TPS as the natural way to represent their many-body state. As a by-product of our investigation, we propose the interpretation of the KW operation as the dipolar version of the discrete Fourier transformation and give an intuitive explanation for the non-invertibility of the KW operation. To keep the main text focused, we collect the relevant results concerning the KW operators in App.~\ref{sec:KW}.

\section{$\mathbb{Z}_N$ cluster state}
\label{sec:MPS-review}

We begin with a self-contained review of the $\Z_N$ cluster state (CS) as the paradigmatic example of SPT phase protected by two global symmetries pertaining to the conservation of $\Z_N$ charges on even and odd sublattices, respectively. The wavefunction of the CS state represents pairwise interaction between two adjacent $\Z_N$ spins. Such interacting wavefunction is  rendered into what we call the $P$-representation involving the product of local projector $P^s$ and the interaction matrix $\bm \Omega$. The matrix $\bm \Omega$ in turn  can be decomposed as a product of two weight matrices, e.g. ${\bm \Omega} = \W \W^t$, with the matrix $\W$ encoding the interaction between physical (visible) and virtual (hidden) variables. Depending on how the $\W$ matrices are split around each local projector, one ends up with different versions of local MPS matrices all leading to the same wavefunction. 

\subsection{Definition and basic properties} 

We begin by reviewing basic properties of $\Z_N$ cluster state. The CS Hamiltonian 
\begin{align}
H_c = -\sum_{j=1}^{L/2} ( Z_{2j-2}^\dag X_{2j-1} Z_{2j} + Z_{2j-1} X_{2j} Z^\dag_{2j+1} + {\rm H.c.})  
\label{cluster-H} 
\end{align}
is defined on a closed chain of even length $L$ in terms of generalized Pauli operators $X,Z$ acting on the $\mathbb{Z}_N$ basis $|s\rangle$ as 
\begin{align} Z|s\rangle = \omega^s |s\rangle , ~~ X |s\rangle = |s+1\rangle . 
\end{align} 
The model has two $\mathbb{Z}_N$ global symmetries~\cite{han24}:
\begin{align} C_1 = \prod_j X_{2j-1} , ~~ C_2 = \prod_j X_{2j} ,
\end{align}
representing the sum of $\Z_N$ charges on the odd and even sublattice, respectively.
Its unique ground state $|\psi\rangle = \sum_{\bf s} \Psi (\mathbf{s}) |{\bf s}\rangle$, ${\bf s} = \{ s_1 , \cdots , s_L \}$, has the wavefunction
\begin{align}
    \Psi({\bf s}) = \omega^{s_1 s_2 - s_2 s_3 + \cdots - s_L s_1 } 
    \label{CS-wf-original} 
\end{align}
with $\omega = \exp (2\pi i /N)$. The invariance of $\Psi ({\bf s})$ under the $C_1$ symmetry operation $s_{2j-1} \rightarrow s_{2j-1} - 1$ or the $C_2$ operation $s_{2j} \rightarrow s_{2j} - 1$ is readily checked. 

The same wavefunction $\Psi({\bf s})$ can be cast as MPS:
\begin{align} 
\Psi ({\bf s}) = {\rm Tr}[ A^{s_1} \cdots A^{s_L} ], 
\label{CS-as-MPS} 
\end{align}
provided the local tensors are chosen as
\begin{align} 
A^{s_{2j-1}} & = \sum_{g \in \Z_N}  \omega^{-s_{2j-1} g} |g\rangle \langle s_{2j-1} |\,, \nn 
A^{s_{2j}} & = \sum_{g \in \Z_N} \omega^{s_{2j} g} |g\rangle \langle s_{2j} | \,.
\label{cluster-MPS0}
\end{align}
Invariance of $\Psi ({\bf s})$ under $C_1^\dagger, C_2^\dagger$ can be checked in its MPS representation by observing that changing $s\rightarrow s+1$ results in 
\begin{align}
\label{how-A-transforms-in-CS} 
A^{s_{2j-1}+1} & = Z^\dag A^{s_{2j-1}} X^\dag, \nn 
A^{s_{2j}+1} & = Z A^{s_{2j}} X^\dag , 
\end{align}
in a manifestation of the push-through condition of MPS. The wavefunction itself undergoes the change 
\begin{align} 
\Psi ({\bf s}) & \xrightarrow{C_1^\dagger} {\rm Tr}[\cdots A^{s_{2j-1}+1} A^{s_{2j}} A^{s_{2j+1}+1} \cdots ] \nn
&= {\rm Tr} [ \cdots ( Z^\dag A^{s_{2j-1}} X^\dag )  A^{s_{2j}} ( Z^\dag A^{s_{2j+1}} X^\dag )  \cdots ], 
\end{align} 
where we used \eqref{how-A-transforms-in-CS} in the second line. The various virtual $X$ operations in the second line can be re-grouped as
\begin{align} 
& {\rm Tr}[ \cdots Z^\dag )  A^{s_{2j-1}} ( X^\dag  A^{s_{2j}} Z^\dag )  A^{s_{2j+1}} ( X^\dag A^{s_{2j+2}} Z^\dag )   \cdots ] .  \nonumber 
\end{align}
The identity $X^\dag A^{s_{2j}} Z^\dag = A^{s_{2j}}$ obeyed by even-site MPS matrices ensures that this is equal to the original CS wavefunction. Another way to see the restoration of the wavefunction is to note that, taking a pair of matrices $A^{s_{2j}} A^{s_{2j+1}}$ as a unit, the transformation under $C_1^\dagger$ gives
\begin{align} 
A^{s_{2j}} A^{s_{2j+1}}\xrightarrow{C_1^\dagger}  A^{s_{2j}} Z^\dag A^{s_{2j+1}} X^\dag = X ( A^{s_{2j}} A^{s_{2j+1}} ) X^\dag . 
\end{align} 
The auxiliary matrices $X$ and $X^\dag$ cancel each other out in taking the product over all sites. A similar exercise using the identity $X^\dag A^{s_{2j-1}} Z = A^{s_{2j-1}}$ for the odd-site matrices shows that the wavefunction is preserved under $C_2^\dagger$ as well.

Next we turn to the open chain of even length $L$, and consider the CS Hamiltonian: 
\begin{align}
    H =& - Z_{2}^\dag X_{3} Z_{4} - Z_{1} X_{2} Z^\dag_{3}  - \cdots \nn
    &- Z_{L-2}^\dag X_{L-1} Z_{L} - Z_{L-3} X_{L-2} Z^\dag_{L-1} + {\rm H.c.} . 
\label{cluster-H_open}
\end{align}
Ground states of this model are
\begin{align}
|\Psi(s_1,s_L)\rangle = \sum_{{\bf s}'}\langle s_1 | A^{s_2} \cdots A^{s_{L}} |s_L\rangle |{\bf s}'\rangle , 
\label{CS_open}
\end{align}
with the ${\bf s}' = \{s_2 , \cdots s_{L-1}\}$ and fixed $\{s_1 , s_L \}$. For this state we show in App.~\ref{app:fractionalization-of-cluster-state}
\begin{align}
C_1^\dagger |\Psi (s_1 , s_L) \rangle & =  X_1^\dag Z_2^\dag Z_L |\Psi (s_1 , s_L)\rangle\,, \nn 
C_2^\dagger |\Psi (s_1 , s_L)\rangle & = Z_1 Z_{L-1}^\dag X_L^\dag |\Psi (s_1 , s_L)\rangle \,,
\label{symm-frac-in-CS} 
\end{align}
demonstrating how the global symmetries $C_1^\dagger$ and $C_2^\dagger$ fractionalize into edge operators $L_i,R_i,\,i=1,2$:
\begin{align}
    C_1^\dagger & \mapsto L_1R_1,\quad (L_1 , R_1 ) = ( X_1^\dag Z_2^\dag, Z_L ) , \nn
    C_2^\dagger & \mapsto L_2R_2,\quad ( L_2 , R_2 ) = ( Z_1, Z_{L-1}^\dag X_L^\dag ) .
\label{sym-frac}
\end{align}
While $C_1^\dagger , C_2^\dagger$ commute, their fractionalizations commute only projectively: 
\begin{align} 
L_1L_2 = \omega L_2L_1, \quad R_2R_1= \omega R_1R_2 , 
\end{align} 
generating the $N$-fold degeneracy at each edge. The corresponding eigenstates can be constructed explicitly as, e.g. $|\Psi(s_1,s_L)\rangle = L_1^{s_1}R_2^{s_L}|\Psi(0,0)\rangle$, using $\{s_1 ,s_L\}$ as quantum numbers labeling each ground state. 

\subsection{CS wavefunction in $P$, NQS, and MPS representations}

We begin with a simple observation that the CS wavefunction becomes
\begin{align} 
\Psi ({\bf s} ) = {\rm Tr} [ P^{s_1} {\bm \Omega}  P^{s_2} {\bm \Omega}^* P^{s_3} {\bm \Omega} \cdots P^{s_L} {\bm \Omega}^* ] ,  \label{psiOmega} 
\end{align}
where $P^s = |s\rangle \langle s|$ is the projector and the interaction matrix $\bm \Omega$ is
\begin{align}
\label{interaction-matrix-Omega} 
{\bm \Omega} & = \sum_{g_1 , g_2 \in \Z_N} \omega^{g_1 g_2} |g_1 \rangle \langle g_2 | . 
\end{align} 
Note that $g_1, g_2$ are virtual, or bond $\Z_N$ variables. The representation in Eq.~\eqref{psiOmega} is equivalent to the pair-correlator product state formulation~\cite{Changlani09}, where the interaction matrix \(\bm \Omega\) serves as a two-site correlator. In the tensor-network formulation of correlator product state (CPS)~\cite{clark18}, correlators are connected through copy tensors at the physical sites. Our projector \(P^s\) is the corresponding copy tensor with its physical index fixed to \(s\), generalized here to \(\mathbb Z_N\) variables. We retain the phrase `\(P\)-representation' over CPS throughout this work to emphasize the role of the local projectors in the subsequent symmetry analysis.

In turn, the interaction $\omega^{ss'}$ between adjacent spins can be thought of as the effective interaction emerging from the interaction of physical spins $s,s'$ with a common hidden variable $h$. Explicitly, one can look for the RBM representation
\begin{align}\label{ww}
    \sum_{h\in \Z_N} W(s_1,h)W(s_2,h) = \omega^{s_1 s_2} , 
\end{align}
fulfilled by some weight function $W(s,h)$. The explicit construction of $W(s,h)$ for $\Z_2$ CS was found before~\cite{sarma17a,duan17}. Here we provide its $\Z_N$ generalization in closed form:
\begin{align}\label{znW}
    W(s,h) & =\kappa^{-1/2} \omega^{ a h^2 + b s^2+c s h} \nn 
    \kappa & = \sum_{h=0}^{N-1}\omega^{2a h^2} , 
\end{align} 
with the coefficients
\begin{align}
    a = \frac{N-1}{4}\,, \quad b= \frac{N-1}{2}\,, \quad c= - 1\,. 
\end{align}
For even $N$, it suffices to choose $(a,b,c) = (-1/4,-1/2,-1)$. For derivation, see App.~\ref{app:derivation-of-W}. We can write \eqref{znW} as a matrix relation: 
\begin{align} 
\W & = \sum_{g,h \in \Z_N} W(g,h) |g\rangle \langle h |  \nn 
\W^t & = \sum_{g,h \in \Z_N} W(g,h) |h\rangle \langle g | \nn 
{\bm \Omega} & = \W \W^t .
\label{W-and-Omega}
\end{align}
Plugging the last line ${\bm \Omega} = \W \W^t$ into the $P$-representation \eqref{psiOmega} results in the NQS representation of the CS wavefunction:
\begin{align} 
\Psi ({\bf s} ) = {\rm Tr} [ P^{s_1} \W \W^t P^{s_2} \W^* \W^\dag P^{s_3} {\bm \Omega} \cdots P^{s_L} \W^* \W^\dag ] . 
\label{NQS-rep-for-CS} 
\end{align}

Finally, we arrive at the MPS representation of the CS wavefunction by noting that Eq.~\eqref{psiOmega} can be written as a matrix product ${\rm Tr}[A^{s_1} A^{s_2} \cdots ]$ provided we set 
\begin{align}
\label{cluster-MPS-v2} 
A^{s_{2j-1}} = {\bm \Omega}^* P^{s_{2j-1}} , ~~  A^{s_{2j}} & = {\bm \Omega} P^{s_{2j}} .
\end{align}
This is precisely the MPS matrix given earlier in Eq.~\eqref{cluster-MPS0}. There is an alternative way to derive the matrix $A^s$ which starts from the NQS representation in \eqref{NQS-rep-for-CS} and groups terms as   
\begin{align} 
A^{s_{2j-1}} & = \W^\dag P^{s_{2j-1}} \W \,, \nn 
A^{s_{2j}} & = \W^t P^{s_{2j}} \W^*  \,.  
\label{A-in-terms-of-W} 
\end{align}
The matrix elements of these MPS are
\begin{align}
A^{s_{2j-1}}_{\alpha\beta} &= \frac{1}{|\kappa|} \omega^{\frac{1}{4} (\alpha^2 - \beta^2 ) + s_{2j-1} (\alpha - \beta)}\,, \nonumber
\\
A^{s_{2j}}_{\alpha\beta} &= \frac{1}{|\kappa|} \omega^{-\frac{1}{4} (\alpha^2 - \beta^2 ) -s_{2j} (\alpha - \beta)}\,.
\end{align}

We began by writing the CS wavefunction in the $P$-representation as the (trace of) the product of projectors and interaction matrices, then performed decomposition of the interaction matrix as a product of two weight matrices to arrive at its NQS representation. The MPS representation of the CS wavefunction follows by identifying the local MPS matrix with the product of a projector and an interaction matrix. Depending on how the identification goes, we may have more than one way to write the MPS matrix, e.g. Eq.~\eqref{cluster-MPS-v2} and Eq.~\eqref{A-in-terms-of-W}, that are unitarily equivalent. The MPS=NQS correspondence demonstrated here for the $\Z_N$ cluster state is consistent with earlier observations relating NQS and tensor-network states~\cite{cirac18,clark18}.

\subsection{Symmetry transformations}
It has been a common practice to prove the invariance of the CS wavefunction under the global symmetries $C_1 , C_2$, as well as their fractionalization in an open chain, in the framework of MPS representation. Here we show that the same analysis can be performed just as easily, if not more so, using the $P$-representation. 

\begin{figure}[h]
    \centering
    \includegraphics[width=0.8\columnwidth]{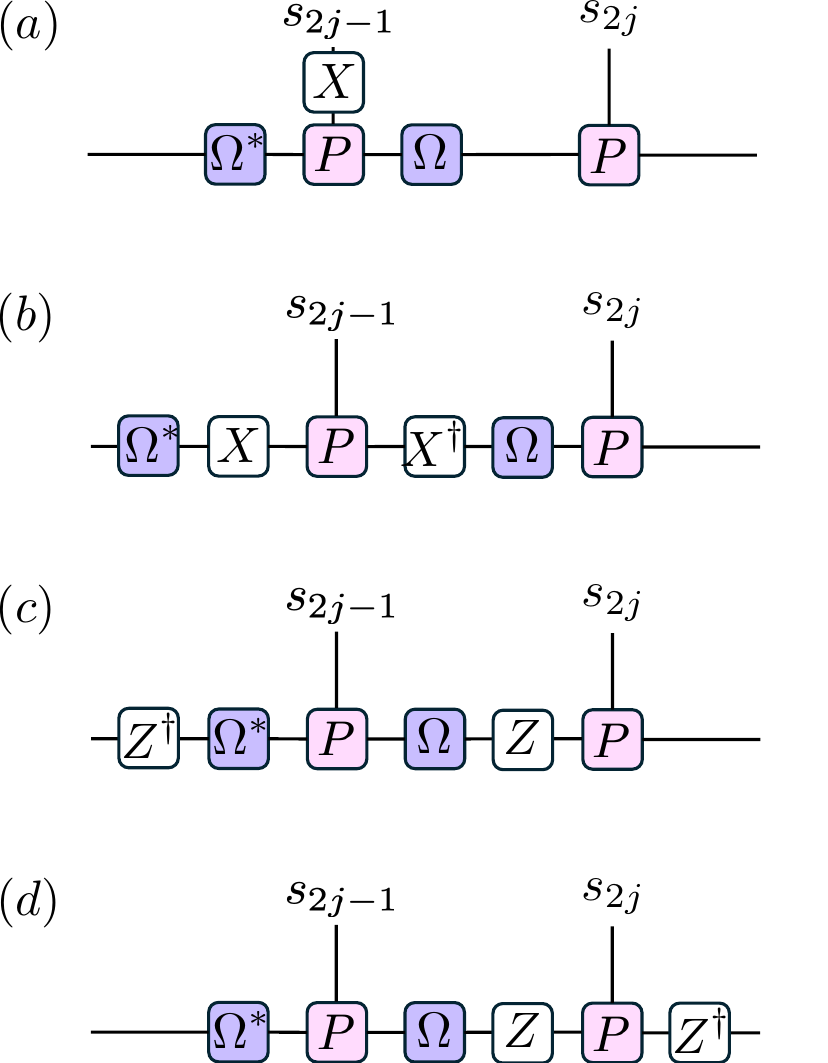}
    \caption{Graphical proof of the invariance of the CS wavefunction under $C_1$ in the $P$-representation. The symmetry operation by $X$ on the projector $P^s$ at the odd sites shown in (a) becomes virtual operations on the same projector as shown in (b). The left (right) action on $\bm \Omega$ ($\bm \Omega^*$) becomes the right (left) action on $\bm \Omega$ ($\bm \Omega^*$) as shown in (c). The net result is the conjugation of the even-site projector by $Z$ and $Z^\dag$ as shown in (d), which gives back the original projector and recovers the CS wavefunction.}
    \label{fig:ZNOmega_sym}
\end{figure}

The symmetry transformation of the projector $P^s \rightarrow P^{s+1}$ results in
\begin{align} P^{s+1} = X P^s X^\dag ,
\end{align} 
which is in fact the push-through condition. Feeding this change into \eqref{psiOmega} gives
\begin{align} 
\Psi(\mathbf{s}) & \xrightarrow{C_1^\dag} \textrm{Tr}\left[( X P^{s_1} X^\dagger ) \bm\Omega P^{s_2} \bm\Omega^* ( X P^{s_3} X^\dagger ) \cdots P^{s_L} \bm\Omega^*\right]  \nn   
&= \textrm{Tr}\left[P^{s_1} ( X^\dagger \bm\Omega P^{s_2} \bm\Omega^* X ) P^{s_3}  \cdots (X^\dag \bm \Omega P^{s_L} \bm\Omega^* X ) \right] 
\label{psiomega-under-C1} 
\end{align}
after some re-grouping of terms. The matrix $\Omega$ has some easily verified properties
\begin{align}
    X^\dagger \bm\Omega &= \bm\Omega Z\,, ~~ 
    \bm\Omega^* X = Z^\dagger\bm\Omega^*\,, \nonumber
    \\
    \bm\Omega X &= Z\bm\Omega \,, ~~ 
    X^\dagger \bm\Omega^*  = \bm\Omega^*Z^\dagger\,,
    \label{how-omega-transforms}
\end{align}
that can be used to convert the expression in the second line of \eqref{psiomega-under-C1} to 
\begin{align}
\textrm{Tr}\left[P^{s_1} \bm\Omega ( Z P^{s_2} Z^\dagger ) \bm\Omega^*  P^{s_3}  \cdots \bm\Omega ( Z P^{s_L} Z^\dagger ) \bm\Omega^*  \right] . 
\end{align} 
Each term inside the parenthesis obeys $Z P^s Z^\dagger = P^s$, establishing the equivalence of the transformed wavefunction to the original one. The symmetry of the wavefunction under $C_2$ can be proven similarly. The algebraic proof given here can be equivalently illustrated as a graphical proof - see Fig.~\ref{fig:ZNOmega_sym}. 

For the open chain of length $L$, the $P$-representation of the CS wavefunction becomes
\begin{align}
    |\Psi (s_1 , s_L) \rangle =  \sum_{{\bf s}'} \langle s_1 | {\bm \Omega}  P^{s_2} {\bm \Omega}^* P^{s_3} {\bm \Omega} \cdots  P^{s_{L-1}} \bm \Omega |s_L \rangle |\mathbf{s}'\rangle ,
\label{OBC}
\end{align}
with a pair of fixed edge spins $\{s_1 , s_L \}$. This is identical to the MPS representation on an open chain given in Eq.~\eqref{CS_open}. The proof of symmetry fractionalization, shown in Eq. \eqref{symm-frac-in-CS}, automatically follows. 

The invariance of the CS state under global symmetries can be checked explicitly in the  RBM-inspired MPS representation, Eq.~\eqref{A-in-terms-of-W}. The proof, reproduced in App.~\ref{app:second-MPS-invariance-proof}, is quite a bit more complicated than what is required to prove the same statement in the $P$-representation. 

\section{Dipolar cluster state}
Dipolar and other multipolar cluster states have been constructed as part of an effort to construct SPT states protected by spatially modulated symmetries. Here we discuss two types of dipolar CS as representative examples of mSPT and discuss their $P$, NQS, and MPS representations. 

\subsection{Type-I dipolar cluster state}
The dipolar cluster state (dCS) was introduced~\cite{han24} as an example of SPT protected by one uniform and one dipole-modulated symmetries 
\begin{align} C = \prod_j X_j , ~~ D= \prod_j ( X_j )^j .
\label{C&D}
\end{align}
An explicit model Hamiltonian can be constructed with these symmetries,
\begin{align} 
H = -\sum_{j=1}^L (Z_{j-1} Z^\dag_j X_j Z^\dag_j Z_{j+1} + h.c. ) ,
\end{align} 
where $L$ must be divisible by $N$ for a closed chain to unambiguously define the dipole symmetry operator $D$. Its unique ground state on a periodic chain is~\cite{han24}
\begin{align}
    \Psi ({\bf s}) & = \omega^{\sum_{j=1}^L s_j (s_{j+1}-s_j )}  = {\rm Tr}[ A^{s_1} \cdots A^{s_L} ] ,  \nn 
    A^s & = \omega^{- s^2} \sum_g \omega^{sg} |g\rangle \langle s| . 
    \label{psi-for-dCSI}
\end{align}
The invariance of this wavefunction under $C$ and $D$ can be readily confirmed. In particular, the $C$-invariance of the MPS wavefunction follows directly from the relation
\begin{align*} A^{s+1} = (ZX) A^s (ZX)^\dag
\end{align*} 
for the matrix $A^s$ in \eqref{psi-for-dCSI}, which can be verified by an explicit calculation. Under the $D$ transformation,
\begin{align} 
{\rm Tr}[A^{s_1} \cdots A^{s_L} ] & \xrightarrow{D} {\rm Tr}[ A^{s_1} ZX A^{s_2} ZX \cdots A^{s_L}(X^\dag Z^\dag)^L ZX]  \nn 
& = (-1)^{L+L/N} {\rm Tr}[ ( A^{s_1} ZX )( A^{s_2} ZX ) \cdots ( A^{s_L} ZX )] \nonumber 
\end{align} 
where we invoked the identity $(X^\dag Z^\dag)^N=(-1)^{N+1}$ in the second line. Further using the identity $ZX A^s = Z A^s Z^\dag$ restores the original wavefunction, up to a phase factor. This line of proof of invariance was first presented in \cite{han24} and is reproduced here for completeness. This dipolar cluster state is referred to as type-I, or dCS$_{\rm I}$, to distinguish it from the type-II dCS state to be defined shortly.

The dCS$_{\rm I}$ wavefunction in the $P$-representation is simply
\begin{align}
\Psi({\bf s}) & = {\rm Tr} [ \tilde{P}^{s_1} {\bm \Omega} \tilde{P}^{s_2} {\bm \Omega} \cdots \tilde{P}^{s_L} {\bm \Omega} ] , \nn 
\tilde{P}^s & = \omega^{-s^2} |s\rangle \langle s| ,
\label{psiOmega-for-dCS}
\end{align}
using the same interaction matrix $\bm \Omega$ previously introduced in \eqref{W-and-Omega}, and the modified projector $\tilde{P}^s$ containing an extra factor $\omega^{-s^2}$. The NQS representation follows from inserting ${\bm \Omega} = \W \W^t$ in the $P$-representation. The MPS representation, in turn, follows from $A^s = \tilde{P}^s {\bm \Omega}$, which reproduces \eqref{psi-for-dCSI}. Alternatively, one can write the MPS matrix as $A^s = \W^t \tilde{P}^s \W$. Explicitly,
\begin{align}
    A^{s}_{\alpha \beta} = \frac{1}{\kappa} \omega^{ -\frac{1}{4}(\alpha^2+\beta^2) - 2 s^2 - s (\alpha+\beta)}\,.
\end{align}

\begin{figure}[h]
    \centering
    \includegraphics[width=0.9\columnwidth]{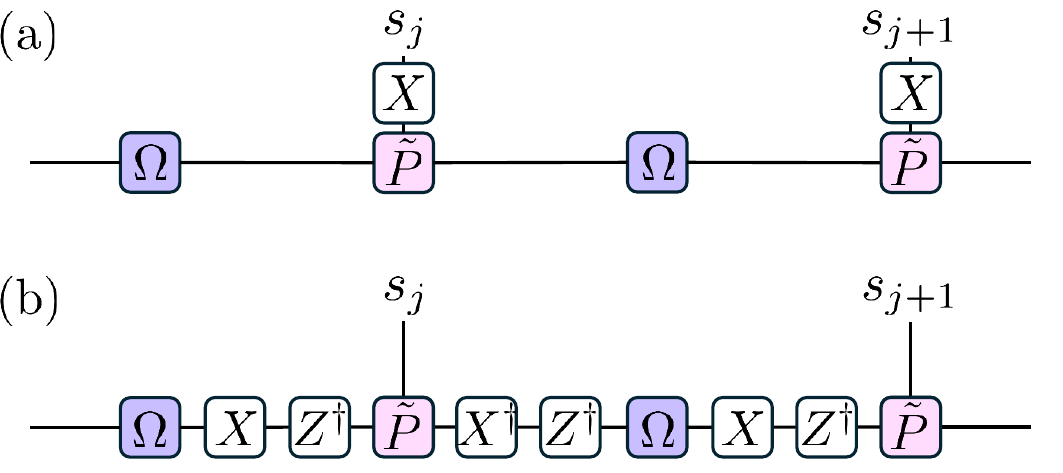}
\caption{Graphical proof of the invariance of the dCS$_{\rm I}$ wavefunction under $C$ in the $P$-representation. Symmetry operation by $X$ on the physical index of each modified projector $\tilde{P}^{s_j}$ shown in a (a) leads to operations on their virtual indices as shown in (b). Viewed as operations on $\bm \Omega$ from the left and the right, they restore the original wavefunction by the push-through condition \eqref{push-Omega-dCSI}. }
    \label{dCSI_C}
\end{figure}

The proof of invariance of the dCS$_{\rm I}$ wavefunction under $C$ and $D$ proceeds by noting that the modified projector $\tilde{P}^s$ obeys the push-through condition 
\begin{align}
    \tilde{P}^{s+1} = Z^\dag X \tilde{P}^s Z^\dag X^\dag = X Z^\dag \tilde{P}^s X^\dag Z^\dag . 
\end{align}
The dCS$_{\rm I}$ wavefunction accordingly transforms as
\begin{align}
    \Psi(\mathbf{s}) \xrightarrow{C^\dagger}& {\rm Tr} [\tilde{P}^{s_1} ( X^\dag Z^\dag \bm\Omega XZ^\dag ) \tilde{P}^{s_2} ( X^\dag Z^\dag \bm\Omega XZ^\dag ) \cdots \nn & \quad \cdots \tilde{P}^{s_L} ( X^\dag Z^\dag \bm\Omega XZ^\dag ) ] ,
\end{align}
where we introduced parentheses to group terms around each $\bm \Omega$. Since $\bm \Omega$ satisfies its own push-through condition
\begin{align}
X^\dag Z^\dag \bm\Omega XZ^\dag= \bm\Omega,
\label{push-Omega-dCSI}
\end{align}
the invariance of the wavefunction $\Psi ({\bf s})$ under $C$ is established. The same proof can be performed step by step in a graphical manner, as illustrated in Fig.~\ref{dCSI_C}. 

\begin{figure}[h]
    \centering
    \includegraphics[width=0.9\columnwidth]{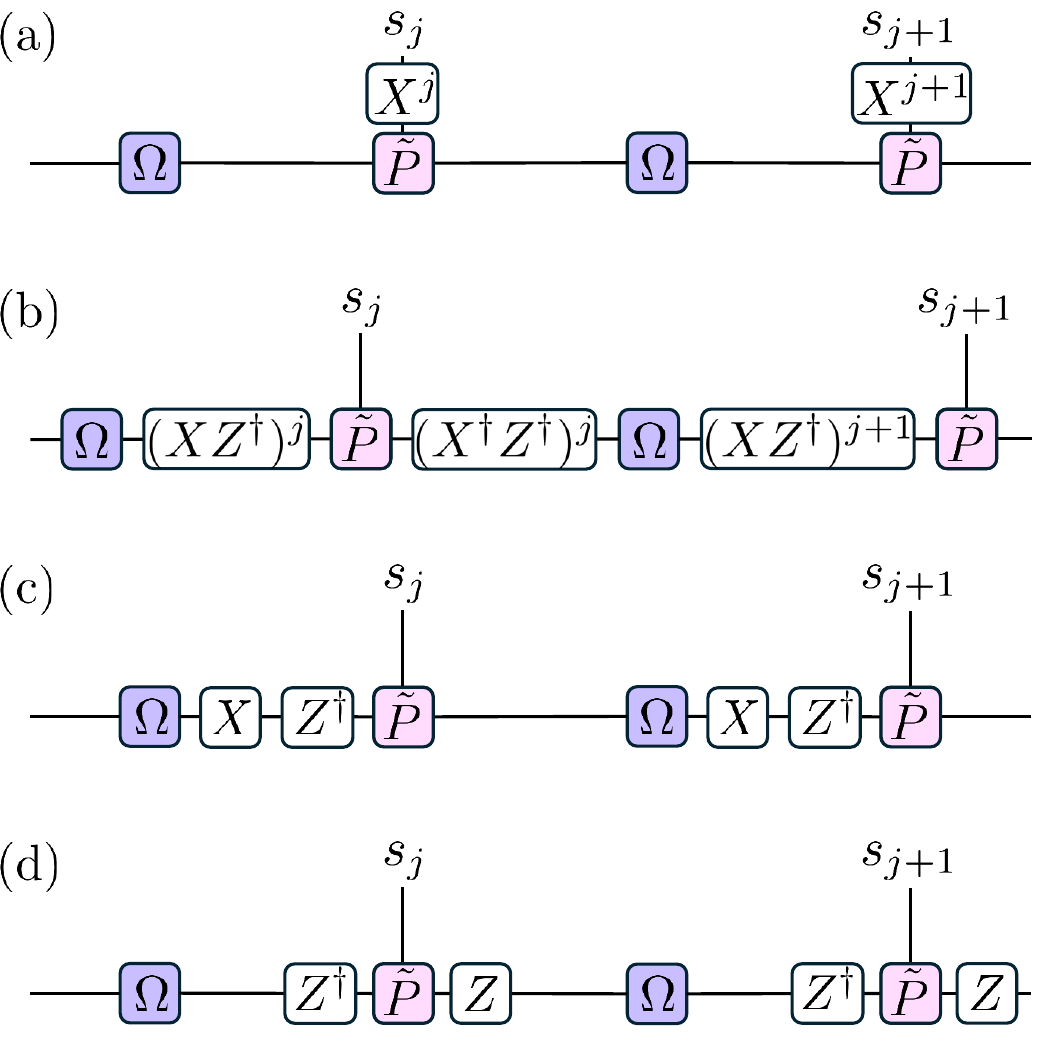}
\caption{Graphical proof of the invariance of the dCS$_{\rm I}$ wavefunction under $D$ in the $P$-representation. Symmetry operation by $X^j$ on the physical index of each modified projector $\tilde{P}^{s_j}$ shown in (a) leads to operations on their virtual indices as shown in (b). The resulting structure can be viewed as operations on $\bm \Omega$ from the left and the right. Exploiting the push-through condition \eqref{push-Omega-dCSI} repeatedly, most of the terms cancel out except for one factor of $XZ^\dag$ on the right side of each $\bm \Omega$ as shown in (c). Finally, the $X$ on the right side of $\bm \Omega$ transforms to $Z$ on its left according to \eqref{how-omega-transforms}. The only remaining term is $Z$ conjugating each modified projector $\tilde{P}^{s_j}$ as shown in (d), which restores the original wavefunction since $Z^\dag \tilde{P}^s Z=\tilde{P}^s$.}
    \label{dCSI_D}
\end{figure}

The proof of invariance under $D$ proceeds by first noting the push-through condition
\begin{align}
    \tilde{P}^{s_j +j } = ( Z^\dag X )^j \tilde{P}^{s_j}  ( Z^\dag X^\dag )^j  = ( X Z^\dag )^j \tilde{P}^{s_j} ( X^\dag Z^\dag )^j . 
\label{push-P-dCSI}
\end{align}
Rather than going through the rest of the derivation algebraically, we refer to the graphical proof in Fig.~\ref{dCSI_D}. That argument reproduces the result $\Psi(\mathbf{s}) \xrightarrow{D^\dagger} (-1)^{L+L/N}\, \Psi(\mathbf{s})$ from the MPS representation up to the overall sign; The correct sign is then recovered by taking into account $(X^\dag Z^\dag)^L=(-1)^{L+L/N}$. Other aspects of the dCS$_{\rm I}$ such as the symmetry fractionalization can be worked out by going through similar analyses as before.

\subsection{Type-II dipolar cluster state: $P$-, NQS, and TPS representations}
A second type of dipolar cluster model was proposed in \cite{han25} with the Hamiltonian  
\begin{align} 
H = - \sum_j Z_{2j-1} X_{2j} Z_{2j+1}^{-2} Z_{2j+3} - \sum_j Z_{2j-2} Z_{2j}^{-2} X_{2j+1}Z_{2j+2} + h.c. ,
\label{type-II-dCS-H} 
\end{align} 
which is symmetric under two sets of charge and dipole symmetries 
\begin{align} 
C_1 &= \prod_j X_{2j-1} , ~~ C_2 = \prod_j X_{2j} \nn 
D_1 &= \prod_j ( X_{2j-1} )^j , ~~ D_2  = \prod_j ( X_{2j} )^j . 
\end{align} 
To distinguish this from the previous dipolar CS, we refer to the new model as dCS$_{\rm II}$. The ground-state wavefunction of the dCS$_{\rm II}$ Hamiltonian on a closed chain is~\cite{han25}:
\begin{align}
\Psi ({\bf s}) 
& = \prod_j\omega^{s_{2j}(s_{2j-1} - 2s_{2j+1} + s_{2j+3})} \nn 
& =\prod_j \Omega_{s_{2j-1}, s_{2j}} \tilde{\Omega}_{s_{2j}, s_{2j+1}} \Omega_{s_{2j},s_{2j+3}}\, ,
\label{dZNRBM}
\end{align}
where $\Omega_{ss'} = \omega^{ss'}$ and $\tilde{\Omega}_{ss'} = \omega^{-2ss'}$. 

\begin{figure}[h]
    \centering
    \includegraphics[width=0.9\columnwidth]{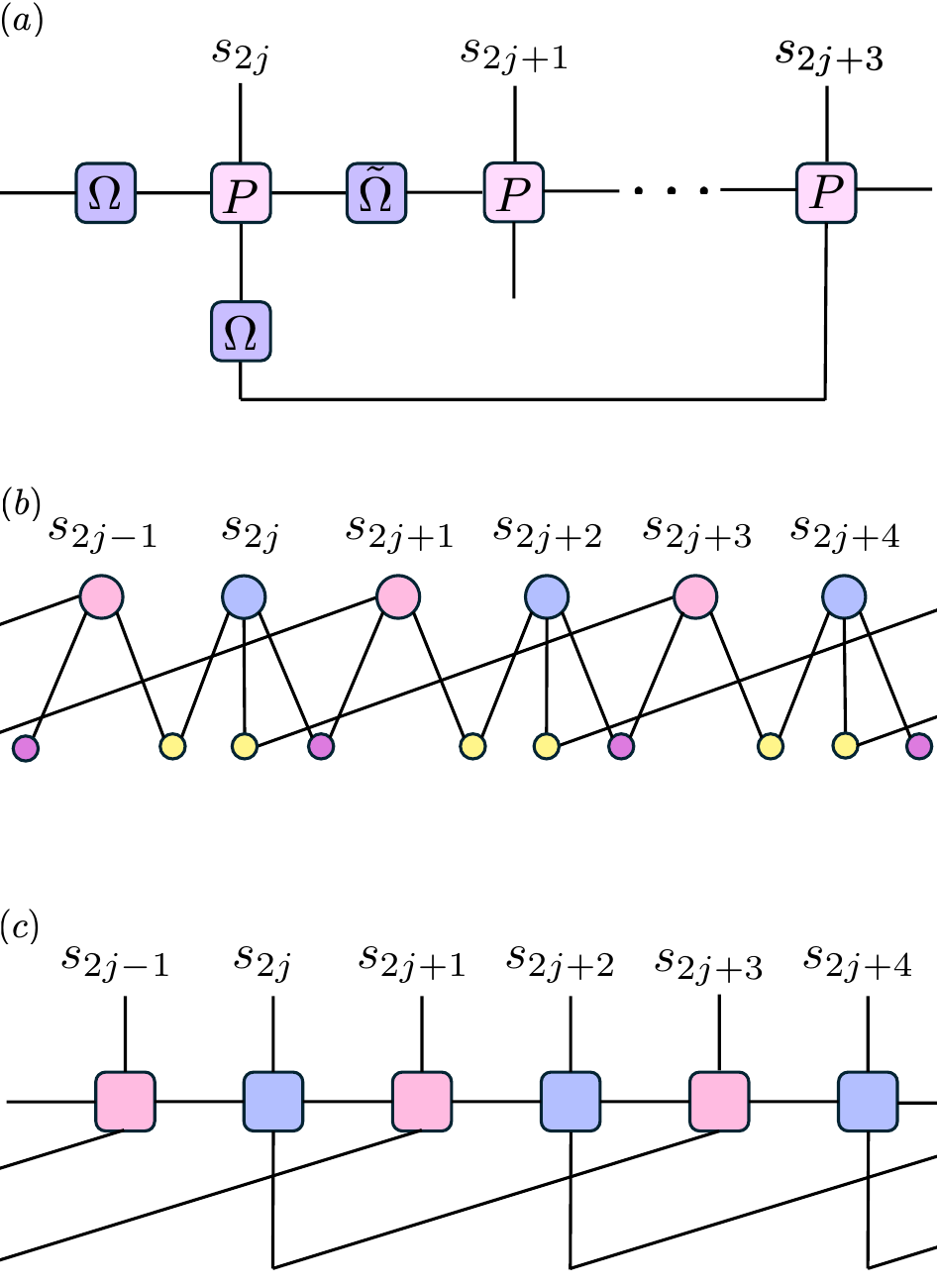}
    \caption{(a) $P$-representation of the dCS$_{\rm II}$ wavefunction. (b) NQS representation of the dCS$_{\rm II}$ wavefunction. Blue and pink circles indicate visible variables on the even and odd sites respectively. The yellow and purple dots represent connections between hidden and visible units by $W$ and $\tilde{W}$ respectively. (c) TPS representation of the dCS$_{\rm II}$ wavefunction. Blue and pink squares indicate the tensors $T^{s_{2j}}$ and $T^{s_{2j+1}}$ respectively. Each tensor has one physical and three virtual bonds.}
    \label{fig:allreps}
\end{figure}

To cast the wavefunction in the $P$-representation, we introduce the projector $P^s$ with three virtual indices~\footnote{In general, we can define the projector with $n$ virtual legs as $P^s_{a_1 a_2 \cdots a_n} = \delta_{s a_1} \delta_{s a_2} \cdots \delta_{s a_n}$},
\begin{align}\label{projector}
P^s_{\alpha\beta\gamma} = \delta_{s\alpha}\delta_{s\beta}\delta_{s\gamma} . 
\end{align} 
The dCS$_{\rm II}$ wavefunction then becomes
\begin{align}
    \Psi(\mathbf{s}) &=\sum_{\bm{\alpha}\bm{\beta}\bm{\gamma} }\sum_{\bm{\alpha}'\bm{\beta}'\bm{\gamma}'}\prod_j P^{s_{2j}}_{\alpha'_j\beta'_j\gamma'_j}\Omega_{\alpha'_j\alpha_j}\tilde{\Omega}_{\beta'_j\beta_j}\Omega_{\gamma'_j\gamma_j}P^{s_{2j+1}}_{\alpha_{j+1}\beta_j\gamma_{j-1}}\,,
    \label{dCS2-using-projectors} 
\end{align}
with ${\bm \alpha} = \{\alpha_1 , \cdots \alpha_L \}$, etc.. It is illustrated graphically in Fig.~\ref{fig:allreps} (a). By appropriately organizing terms around each site, we can directly obtain a TPS representation with the local tensors carrying three virtual indices: 
\begin{align}
    \Psi(\mathbf{s}) &=\sum_{\bm{\alpha}\bm{\beta}\bm{\gamma} }\prod_j T^{s_{2j}}_{\alpha_j \beta_j \gamma_j} T^{s_{2j+1}}_{\alpha_{j+1}\beta_j \gamma_{j-1}}\,, \nonumber
    \\
    T^{s_{2j}}_{\alpha_j \beta_j \gamma_j} &= \sum_{\alpha'_j\gamma'_j} P^{s_{2j}}_{\alpha'_j\beta_j\gamma'_j}\Omega_{\alpha'_j\alpha_j}\Omega_{\gamma'_j\gamma_j} \,, \nonumber
    \\
    T^{s_{2j+1}}_{\alpha_{j+1}\beta_j \gamma_{j-1}} &= \sum_{\beta'_j}P^{s_{2j+1}}_{\alpha_{j+1}\beta'_j\gamma_{j-1}} \tilde{\Omega}_{\beta_j \beta'_j} \,.
\label{dCSII-TPS-P}
\end{align}

To cast the wavefunction in the NQS representation, we introduce a second weight function $\tilde{W}(s,h)$ satisfying
\begin{align}
    \sum_h \tilde{W}(s_1 ,h)\tilde{W}(s_2 ,h) &= \omega^{-2s_1 s_2}\,, \label{tildeW}
\end{align}
in addition to $W(s,h)$ given in \eqref{ww}. Such function can be found
\begin{align}
    \tilde{W}(s,h) & = \tilde{\kappa}^{-1/2} \omega^{\tilde ah^2 + \tilde b s^2 + \tilde csh} \nn  
    \tilde{\kappa} & = \sum_{h=0}^{N-1} \omega^{2\tilde ah^2} 
\end{align}
with
\begin{align}
    (\tilde a,\tilde b,\tilde c) = \left(\frac{1}{2}, 1, 2\right)\, ,
\end{align}
for $N \neq2\mod4$. For $N = 2 \mod4$, the constants are modified to
\begin{align}
    (\tilde a,\tilde b,\tilde c) = \left(\frac{N+2}{4}, 1 +\frac{N}{2}, 2\right)\,.
\end{align}
For derivation of the coefficients, see App.~\ref{app:derivation-of-W}. Plugging the $W$'s and $\tilde{W}$'s into \eqref{dZNRBM} gives the NQS representation of the dCS$_{\rm II}$ wavefunction. It is shown graphically in Fig.~\ref{fig:allreps} (b). 

One can group the various $W$ and $\tilde{W}$ weights around each site and arrive at an alternative TPS representation: 
\begin{align}\label{dZNTN}
    \Psi(\mathbf{s}) 
    =\sum_{\bm{\alpha}\bm{\beta}\bm{\gamma} }\prod_j T^{s_{2j}}_{\alpha_j\beta_j\gamma_j}T^{s_{2j+1}}_{\alpha_{j+1}\beta_j\gamma_{j-1}} , 
\end{align}
where
\begin{align}
\label{dCSII-TPS-W}
T^{s_{2j}}_{\alpha_j\beta_j \gamma_j} & = W(s_{2j},\alpha_j)\tilde{W} (s_{2j},\beta_j)W(s_{2j},\gamma_j) \nn 
& = \frac{1}{\sqrt{\kappa^2 \tilde{\kappa}}}\omega^{-\frac{1}{4}(\alpha_j^2 + \gamma_j^2 -2\beta_j^2) - s_{2j}(\alpha_j + \gamma_j -2\beta_j)} \nn 
T^{s_{2j+1}}_{\alpha_{j+1}\beta_j \gamma_{j-1}} &=W(s_{2j+1},\alpha_{j+1})\tilde{W}(s_{2j+1},\beta_j)W(s_{2j+1},\gamma_{j-1}) \nn 
&= \frac{1}{\sqrt{\kappa^2 \tilde{\kappa}}}\omega^{-\frac{1}{4}(\alpha_{j+1}^2 + \gamma_{j-1}^2 -2\beta_j^2) -s_{2j+1}(\alpha_{j+1} + \gamma_{j-1} -2\beta_j)} . 
\end{align}

\begin{figure}[h]
    \centering
    \includegraphics[width=0.9\columnwidth]{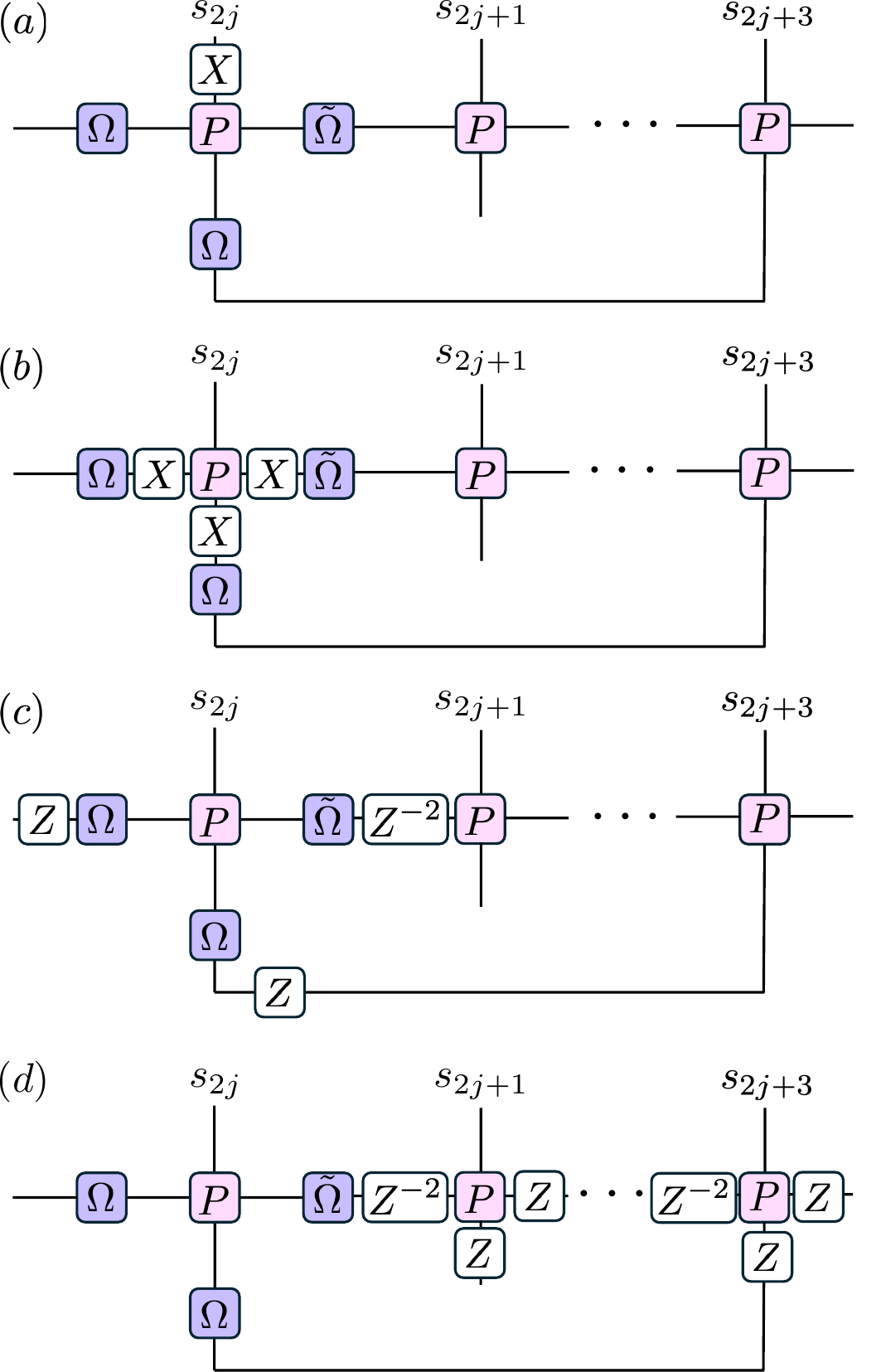}
\caption{Graphical proof of the invariance of the dCS$_{\rm II}$ wavefunction under $C_2$ in the $P$-representation. The symmetry operation by $X$ on the projector $P^s$ at the even sites shown in (a) becomes virtual operations on the same projector as shown in (b). The virtual $X$ operations are moved through the interaction matrices $\bm\Omega$, $\tilde{\bm\Omega}$, which convert them into $Z$ operations as shown in (c). The resultant operations are $Z$ operations around the projectors at odd sites whose sum of exponents is zero as shown in (d), thereby giving the original wavefunction.}
    \label{dZNOmega_sym}
\end{figure}

\begin{figure}[h]
    \centering
    \includegraphics[width=0.9\columnwidth]{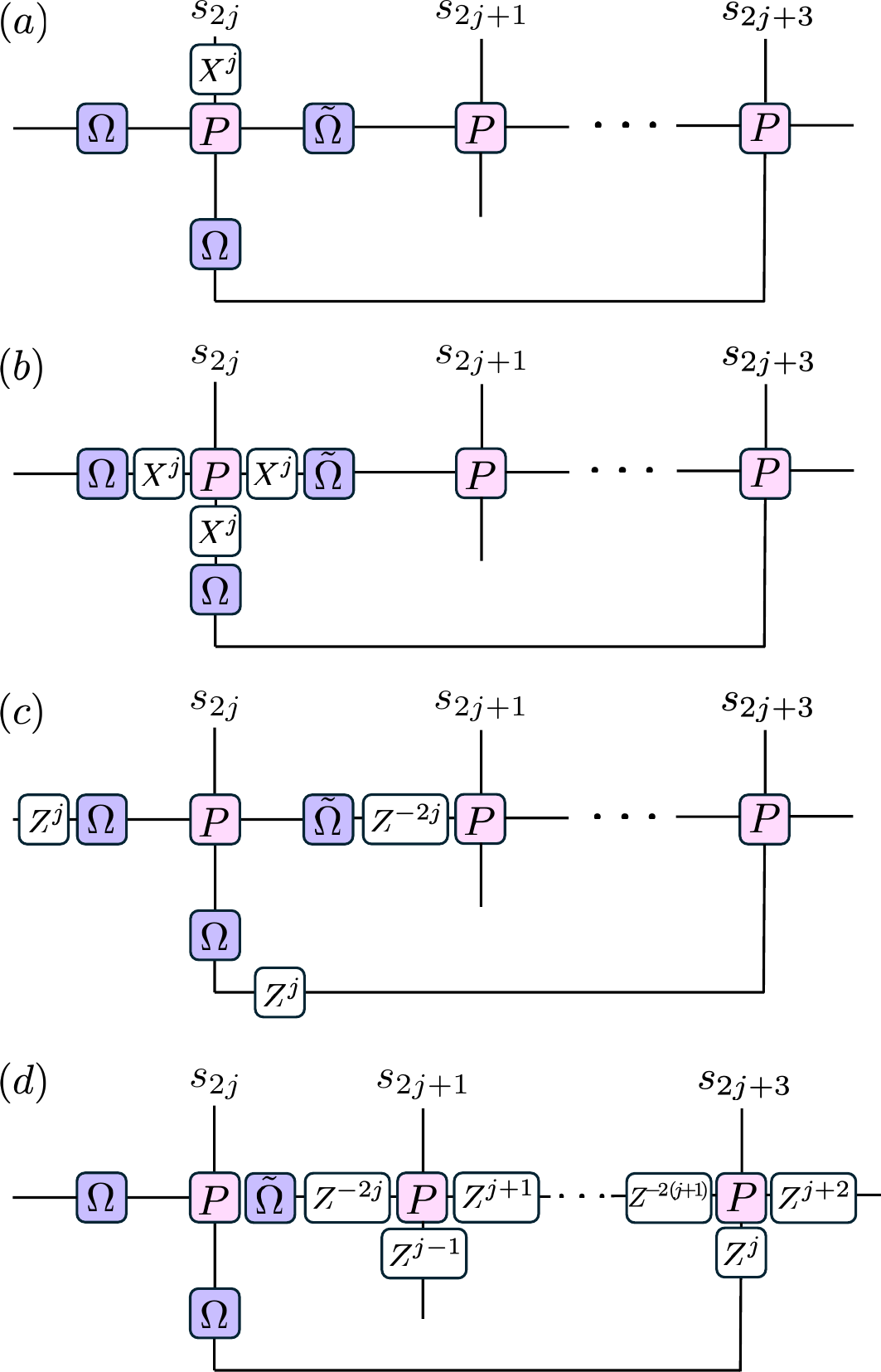}
\caption{Graphical proof of the invariance of the dCS$_{\rm II}$ wavefunction under $D_2$ in the $P$-representation. The symmetry operation by $X^j$ on the projector $P^s$ at the even sites shown in (a) becomes virtual operations on the same projector as shown in (b). The virtual $X^j$ operations are moved through the interaction matrices $\bm\Omega$, $\tilde{\bm\Omega}$, which convert them into $Z$ operations as shown in (c). The resultant operations are $Z$ operations around the projectors at odd sites whose sum of exponents is zero as shown in (d), thereby giving the original wavefunction.}
    \label{dZNOmega_sym_D2}
\end{figure}

Let us now show that the dCS$_{\rm II}$ wavefunction is invariant under all four symmetry operations $C_1, C_2$ and $D_1, D_2$, using its $P$-representation in \eqref{dCS2-using-projectors}. As before, such proof begins by expressing how the physical action $s\rightarrow s+1$ in the projector $P^s_{\alpha\beta\gamma}$ is transferred to the virtual degrees of freedom. The proof is most easily demonstrated graphically, as illustrated in Figs.~\ref{dZNOmega_sym} and \ref{dZNOmega_sym_D2}. The algebraic proof, equivalent to the graphical one, can be found in App.~\ref{app:symmetry_dCS2}.

A proof of symmetry based on the NQS representation of the dCS$_{\rm II}$ wavefunction is possible, by leveraging some transformation properties of the weight functions ($k \in \Z_N$): 
\begin{align}\label{Wshift} 
    W(s+k,h) &= W(s,h) \omega^{2bks+ckh+bk^2}\,, \nn 
    \sum_h W(x,h) W(y,h) \omega^{ckh} &= \omega^{-bk^2} \omega^{-2bxy} \omega^{-2bk(x+y)}\, . 
\end{align}
The $W$ and $\tilde{W}$ weights used in the NQS construction of the dCS$_{\rm II}$ both belong to this family of functions with specific choices of constants $(a,b,c)$. As detailed in App.~\ref{app:symmetry_dCS2}, these relations are sufficient to prove 
\begin{align}\label{C1_dCS}
    \Psi(\mathbf{s}) \xrightarrow[]{C_1, C_2 , D_1 , D_2}  \Psi(\mathbf{s})\,.
\end{align}

A general lesson may be drawn from our endeavor. Modulated SPT states as exemplified by the dCS$_{\rm II}$ state have wavefunctions that reflect interactions among spins going beyond the nearest neighbor. As long as the interactions are of pairwise nature, though, the wavefunction can be readily cast in the $P$-representation, which in turn facilitates the analysis of symmetry invariance and its fractionalization through push-through conditions of the $P$-tensor. Furthermore, the $P$-representation generalizes easily to projectors having multiple virtual legs, resulting in the TPS representation of the wavefunction. Other mSPTs such as those protected by quadrupolar and exponential symmetries~\cite{han24,pace26,han25} are amenable to similar $P$-representation analysis, though we do not pursue the issues here.  

\subsection{Representational aspects of NQS and TPS}
So far our work focused on demonstrating the equivalence of MPS to the NQS representation of the same state embodying various symmetry-protected orders. This has been particularly transparent in the case of cluster state and the type-I cluster state, where the NQS representations give a nice behind-the-scenes interpretation of where the MPS representation with its hidden units come from. In short, the NQS leads to MPS by re-organizing the weight functions around a site rather than around a bond. 

For the type-II dCS state, the MPS-NQS correspondence becomes more intricate, yet the NQS representation remains straightforward to obtain. Specifically, the weight functions $W$ and $\tilde{W}$ can be derived directly from the pairwise structure of the wavefunction in Eq.\,\eqref{dZNRBM}, and the re-organization of these weight functions around each site naturally gives rise to the TPS with rank-3 tensors in Eq.\,\eqref{dCSII-TPS-W}. Arriving at such a tensor network ansatz from the bare wavefunction alone would be considerably less obvious. We believe this conceptual directness — the ability to systematically construct the appropriate tensor network structure starting from the NQS — is an important advantage of the NQS framework for states with modulated symmetries.

\begin{figure}[t]
\includegraphics[width=0.8\columnwidth]{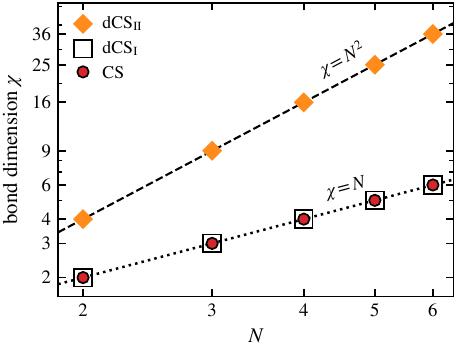}
\caption{Bond dimension $\chi$ of the MPS ground state obtained by DMRG for the $\mathbb{Z}_N$ cluster state (CS), dCS$_{\rm I}$, and dCS$_{\rm II}$ on an open chain of length $L = 24$ with fixed edge spins. The dashed (dotted) line indicates $\chi = N^2$ ($\chi = N$).}
\label{fig:bond_dimension}
\end{figure}

\begin{figure}[t]
\includegraphics[width=\columnwidth]{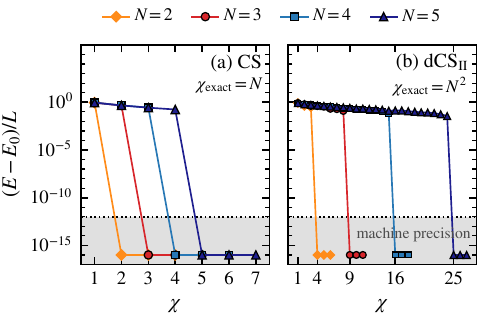}
\caption{Energy per site $(E - E_0)/L$ of the DMRG ground state with the bond dimension restricted to $\chi$, for (a) the $\mathbb{Z}_N$ cluster state and (b) dCS$_{\rm II}$ on an open chain of length $L = 24$. For each $\chi$, the lowest energy out of three random initial MPS is shown. The shaded region indicates $(E - E_0)/L < 10^{-12}$. The exact ground state is reached at $\chi = N$ in (a) and at $\chi = N^2$ in (b).}
\label{fig:chi_scan}
\end{figure}

To compare the TPS with the conventional MPS, we employ DMRG to numerically obtain the MPS ground state of the dCS$_{\rm II}$ Hamiltonian in Eq.~\eqref{type-II-dCS-H} on an open chain of even length $L$. On the open chain we keep only the terms of Eq.~\eqref{type-II-dCS-H} that lie within the chain. No $X$ operator then acts on the four edge spins $s_1, s_3, s_{L-2}, s_L$, which generates the $N^2$-fold degeneracy at each edge. To single out a unique ground state, we add $-(Z_k + Z_k^\dagger)$ for $k = 1, 3, L-2, L$. These terms commute with every term of the Hamiltonian and fix the edge spins to $s_k = 0$, in the same way as the pair of fixed edge spins in Eq.~\eqref{OBC}. Since every term of the resulting Hamiltonian is bounded below by $-2$, and all terms commute, the ground-state energy is $E_0 = -2L$, and the ground state is the dCS$_{\rm II}$ wavefunction of Eq.~\eqref{dZNRBM} with $s_1 = s_3 = s_{L-2} = s_L = 0$.

Each calculation starts from a random MPS with bond dimension $4$ and performs $20$ sweeps with a truncation cutoff of $10^{-12}$. A small noise term is added in the first four sweeps. The maximal bond dimension is set to $2N^2+4$, so that the bond dimension is not restricted. For $L = 24$ and $N = 2, \dots, 6$, the DMRG energy agrees with $E_0$ within machine precision, and the DMRG ground state $|\psi\rangle$ coincides with the exact wavefunction, $1 - |\langle \Psi | \psi \rangle| < 10^{-14}$. The resulting bond dimension of the MPS tensor $A^s$ is $\chi = N^2$ for every $N$, as shown in Fig.~\ref{fig:bond_dimension}. The same calculation for the $\mathbb{Z}_N$ cluster state and the dCS$_{\rm I}$ state gives $\chi = N$. We also performed DMRG with the bond dimension restricted to a given $\chi$. As shown in Fig.~\ref{fig:chi_scan}, the energy reaches $E_0$ only at $\chi = N^2$ for dCS$_{\rm II}$ and at $\chi = N$ for the $\mathbb{Z}_N$ cluster state. One step below, the energy per site remains above $E_0/L$ by $0.04$ to $0.9$.

The bond dimension $\chi = N^2$ follows directly from the pairwise interactions in Eq.~\eqref{dZNRBM}. Consider a cut between the sites $2j$ and $2j+1$. Three interactions connect the two sides of the cut,
$\omega^{s_{2j-2}s_{2j+1}}$, $\omega^{-2s_{2j}s_{2j+1}}$, and $\omega^{s_{2j}s_{2j+3}}$, so that
\begin{align}
\Psi(s) &= \Psi_{\rm left}\, \Psi_{\rm right}\, \omega^{\, x^{t} M y}, \nonumber \\
x &= \begin{pmatrix} s_{2j-2} \\ s_{2j} \end{pmatrix}, \quad
y = \begin{pmatrix} s_{2j+1} \\ s_{2j+3} \end{pmatrix}, \quad
M = \begin{pmatrix} 1 & 0 \\ -2 & 1 \end{pmatrix},
\end{align}
where $\Psi_{\rm left}$ ($\Psi_{\rm right}$) contains only the spins on the left (right) side of the cut. Since $\det M = 1$, the matrix $M$ is invertible for every $N$, and the $N^2 \times N^2$ matrix $\omega^{x^t M y}$ is, up to a factor $N$, the discrete Fourier transform on $\mathbb{Z}_N \times \mathbb{Z}_N$. Its singular value decomposition therefore has $N^2$ nonzero singular values, all equal to each other. For the cut between the sites $2j+1$ and $2j+2$, the two interactions $\omega^{s_{2j}s_{2j+3}}$ and $\omega^{s_{2j+1}s_{2j+2}}$ connect the two sides, and the same argument again gives $N^2$ equal singular values. An MPS of the dCS$_{\rm II}$ ground state thus requires bond dimension $N^2$, and the DMRG ground state realizes this value with all $N^2$ singular values equal. For the $\mathbb{Z}_N$ cluster state and the dCS$_{\rm I}$ state, only a single nearest-neighbor interaction connects the two sides of a cut, and the bond dimension is $N$. It is the interaction with the further neighbor, $\omega^{s_{2j}s_{2j+3}}$, that places two independent $\mathbb{Z}_N$ variables on each side of every cut in the dCS$_{\rm II}$ state. The TPS in Eq.~\eqref{dCSII-TPS-W} carries the interactions across a cut on separate virtual indices of dimension $N$, whereas the MPS must combine them into a single virtual index of dimension $N^2$.

The code and data used for the DMRG calculations are publicly available at \cite{DMRGcode}.


\section{$P$-representation under symmetry-preserving deformations}
\label{sec:deformed}

We now construct deformed Hamiltonians whose exact ground states admit $P$-representations with finite correlation lengths. The construction applies to all three cluster states considered above. These deformations preserve the symmetries of the original cluster Hamiltonians. The $P$-representation continues to provide an exact description of the deformed ground states by adding virtual legs to the projectors and introducing new interaction matrices between distant spins. The construction clearly proves that the $P$-representation offers an intuitive analytical description of the underlying structure of the ground-state wavefunctions under the deformation.

Let $\alpha\in\{\mathrm{CS},\mathrm I,\mathrm{II}\}$ label the three models, and let $K_j^\alpha$ denote their local unitary stabilizers
\begin{align}
K_{2j-1}^{\mathrm{CS}}
&=Z_{2j-2}^\dagger X_{2j-1}Z_{2j}, \nonumber \\
K_{2j}^{\mathrm{CS}}
&=Z_{2j-1}X_{2j}Z_{2j+1}^\dagger, \nonumber\\
K_j^{\mathrm I}
&=Z_{j-1}Z_j^\dagger X_jZ_j^\dagger Z_{j+1}, \nonumber\\
K_{2j}^{\mathrm{II}}
&=Z_{2j-1}X_{2j}Z_{2j+1}^{-2}Z_{2j+3}, \nonumber\\
K_{2j+1}^{\mathrm{II}}
&=Z_{2j-2}Z_{2j}^{-2}X_{2j+1}Z_{2j+2}\,,
\end{align}
satisfying $K_j^\alpha|\psi_0^\alpha\rangle=|\psi_0^\alpha\rangle$ for the corresponding fixed-point ground state.
We write the fixed-point Hamiltonians as
\begin{align}
Q_j^\alpha
&=1-\frac{K_j^\alpha+K_j^{\alpha\dagger}}{2},
\notag\\
H_0^\alpha
&=\sum_j Q_j^\alpha.
\label{eq:deform_fixed_parent}
\end{align}
This only shifts and rescales the Hamiltonians introduced above, without changing their ground states. Since each $K_j^\alpha$ is unitary, $Q_j^\alpha$ is positive semidefinite.

We introduce a separation scale $\ell_\alpha$ for each stabilizer model by 
\begin{equation}
\ell_{\mathrm{CS}}=2,\qquad
\ell_{\mathrm I}=N,\qquad
\ell_{\mathrm{II}}=2N ,
\label{eq:deform_ranges}
\end{equation}
and take $L$ to be a multiple of $\ell_\alpha$ for a periodic chain. The two-site equality projector is defined by 
\begin{equation}
\Pi_{j,k}=\frac{1}{N}\sum_{q=0}^{N-1}(Z_j^\dagger Z_k)^q,
\qquad
\Pi_{j,k}|\mathbf s\rangle=\delta_{s_j,s_k}|\mathbf s\rangle.
\label{eq:deform_projector}
\end{equation}
For a finite deformation parameter $g\geq0$, define the local deformations by
\begin{align}
f_{g,j}^\alpha
&=\exp\!\left[\frac{g}{2}
\bigl(\Pi_{j-\ell_\alpha,j}+
\Pi_{j,j+\ell_\alpha}\bigr)\right],
\label{eq:deform_local_filter}
\end{align}
and the corresponding frustration-free Hamiltonian by
\begin{align} 
H_g^\alpha
&=\sum_j(f_{g,j}^\alpha)^{-1}Q_j^\alpha
(f_{g,j}^\alpha)^{-1}.
\label{eq:deform_parent}
\end{align}
Each term of $H_g^\alpha$ is Hermitian and positive semidefinite. Let $S$ denote any protecting symmetry generator such as $C_1,C_2$ for CS, $C,D$ for dCS$_{\mathrm I}$, and $C_1,C_2,D_1,D_2$ for dCS$_{\mathrm{II}}$. Since each symmetry operator commutes with the projector defined in \eqref{eq:deform_projector}, $[S,\Pi_{j,j+\ell_\alpha}]=0$, the deformed Hamiltonian $H_g^\alpha$ inherits all the symmetries of the undeformed stabilizer model, $[S,H_g^\alpha]=0$. The corresponding ground states of the deformed Hamiltonians are obtained by using the global deformation
\begin{equation}
F_g^\alpha=\exp\!\left(\frac{g}{2}
\sum_{j=1}^{L}\Pi_{j,j+\ell_\alpha}\right),
\label{eq:deform_filter}
\end{equation}
as
\begin{equation}
|\psi_g^\alpha\rangle = F_g^\alpha|\psi_0^\alpha\rangle\,.
\label{eq:deform_state}
\end{equation}
The deformed states have the same symmetry eigenvalues as the corresponding fixed-point states.

To verify that Eq.~\eqref{eq:deform_state} is a ground state, write $F_g^\alpha=f_{g,j}^\alpha G_{g,j}^\alpha$. The diagonal operator $G_{g,j}^\alpha$ contains no factor involving site $j$. Since the only spin shifted by $K_j^\alpha$ is at site $j$, $G_{g,j}^\alpha$ commutes with $K_j^\alpha$ and hence with $Q_j^\alpha$. Therefore,
\begin{equation}
Q_j^\alpha(f_{g,j}^\alpha)^{-1}
F_g^\alpha|\psi_0^\alpha\rangle
=G_{g,j}^\alpha Q_j^\alpha|\psi_0^\alpha\rangle
=0.
\label{eq:deform_groundstate_proof}
\end{equation}
Every term of $H_g^\alpha$ therefore annihilates $|\psi_g^\alpha\rangle$, making it an exact zero-energy ground state. Conversely, if $|\phi\rangle$ has zero energy under $H_g^\alpha$, positivity of the local terms implies $Q_j^\alpha(f_{g,j}^\alpha)^{-1}|\phi\rangle=0$ for every $j$. The same commutation relation then gives $Q_j^\alpha(F_g^\alpha)^{-1}|\phi\rangle=0$. Thus $(F_g^\alpha)^{-1}|\phi\rangle$ is a fixed-point ground state. Since the latter is unique on the periodic chain, the deformed ground state is also unique.  For sufficiently small $g$, stability of the gapped fixed-point Hamiltonians ensures that the symmetry-preserving deformations remain in the corresponding SPT phases.

We can also express the wavefunction Eq.~\eqref{eq:deform_state} as
\begin{align}
\Psi_g^\alpha(\mathbf s)
&=\Psi_0^\alpha(\mathbf s)
\prod_{j=1}^{L}\Lambda(g)_{s_j,s_{j+\ell_\alpha}},
\label{eq:deform_amplitude}\\
\Lambda(g)_{s,t}
&=\exp\!\left(\frac{g}{2}\delta_{s,t}\right).
\label{eq:deform_pair_matrix}
\end{align}
The $N\times N$ interaction matrix $\Lambda(g)$ has diagonal elements $e^{g/2}$ and off-diagonal elements 1. 

\begin{figure}[t]
    \centering
    \includegraphics[width=0.9\columnwidth]{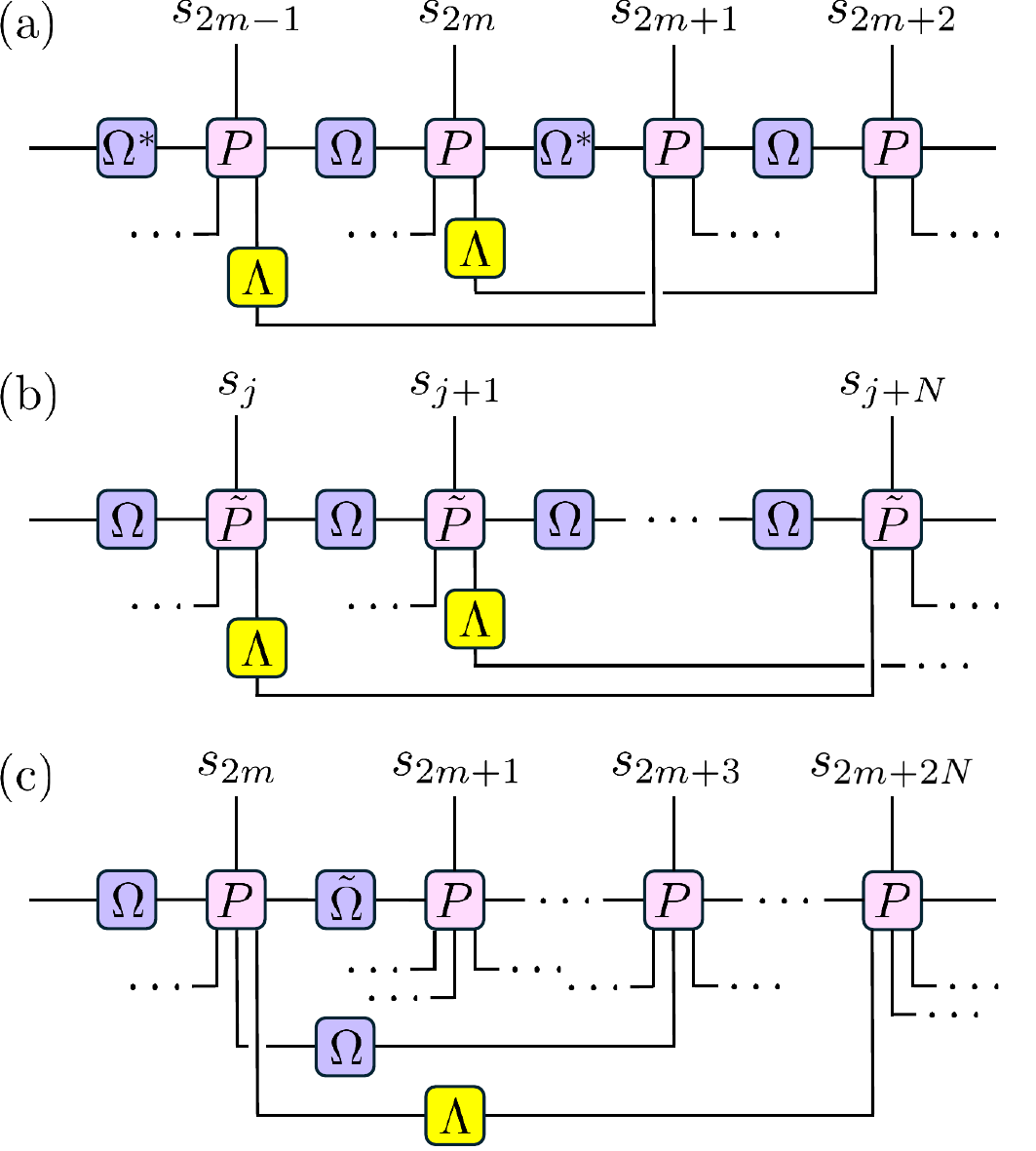}
\caption{$P$-representations of the deformed (a) cluster state, (b) type-I, and (c) type-II dipolar cluster states.}
\label{deformed-P}
\end{figure}

The key observation is that the deformed ground state continues to allow analytic representation within the $P$-representation. Explicitly, we get for the deformed CS
\begin{widetext}
\begin{align}
\Psi_g^{\mathrm{CS}}(\mathbf s)=
\sum_{\mathbf a,\mathbf b,\mathbf u,\mathbf v}
\prod_j\bigg[
P^{s_{2j-1}}_{a_{2j-1}b_{2j-1}u_{2j-1}v_{2j-1}}\Omega_{b_{2j-1}a_{2j}}
P^{s_{2j}}_{a_{2j}b_{2j}u_{2j}v_{2j}}
\Omega^*_{b_{2j}a_{2j+1}}
\Lambda(g)_{u_{2j-1},v_{2j+1}}
\Lambda(g)_{u_{2j},v_{2j+2}}\bigg] . 
\label{eq:deform_prep_cs}
\end{align}
Similarly, the deformed dCS$_{\mathrm I}$ is expressed as
\begin{align}
\Psi_g^{\mathrm I}(\mathbf s)=
\sum_{\mathbf a,\mathbf b,\mathbf u,\mathbf v}
\prod_j\bigg[
\tilde{P}^{s_j}_{a_jb_ju_jv_j}\Omega_{b_ja_{j+1}}
\Lambda(g)_{u_j,v_{j+N}}\bigg].
\label{eq:deform_prep_I}
\end{align}
Finally, the deformed dCS$_{\mathrm{II}}$ is represented as
\begin{align}
\Psi_g^{\mathrm{II}}(\mathbf s) =
\sum_{\mathbf u,\mathbf v}
\sum_{\mathbf a',\mathbf b',\mathbf c'}
\sum_{\mathbf a,\mathbf b,\mathbf c}
\prod_j\bigg[ P^{s_{2j}}_{a'_j b'_j c'_j u_{2j}v_{2j}}
\Omega_{a'_j a_j}
\widetilde\Omega_{b'_j b_j}
\Omega_{c'_j c_j}
P^{s_{2j+1}}_{a_{j+1}b_j c_{j-1}u_{2j+1}v_{2j+1}}
\Lambda(g)_{u_{2j},v_{2j+2N}}
\Lambda(g)_{u_{2j+1},v_{2j+2N+1}}
\bigg].
\label{eq:deform_prep_II}
\end{align}
Each projector of the deformed states acquires two additional virtual legs, each of which is connected to an interaction matrix $\Lambda(g)$. For $g=0$, all $\Lambda(g)_{u,v}$ become one and the additional virtual legs can be removed by summing over the free indices. The $P$-representations are shown graphically in Fig.~\ref{deformed-P}.

Furthermore, we can express the three deformed cluster states in the RBM form 
\begin{align}
\Psi_g^{\mathrm{CS}}(\mathbf s) &=
\sum_{\mathbf a,\mathbf b,\mathbf h}
\prod_j \big[
W(s_{2j-1},a_j)W(s_{2j},a_j)
W^*(s_{2j},b_j)W^*(s_{2j+1},b_j)
\big]
\prod_k W_g(s_k,h_k)W_g(s_{k+2},h_k),
\label{eq:deformed_RBM_CS}
\\
\Psi_g^{\mathrm I}(\mathbf s) &=
\sum_{\mathbf a,\mathbf h}
\prod_j \big[
\omega^{-s_j^2}W(s_j,a_j)W(s_{j+1},a_j)
\big]
\prod_k W_g(s_k,h_k)W_g(s_{k+N},h_k),
\label{eq:deformed_RBM_I}
\\
\Psi_g^{\mathrm{II}}(\mathbf s) &=
\sum_{\mathbf a,\mathbf b,\mathbf c,\mathbf h}
\prod_j \big[
W(s_{2j},a_j)W(s_{2j-1},a_j)
\widetilde W(s_{2j},b_j)\widetilde W(s_{2j+1},b_j)
W(s_{2j},c_j)W(s_{2j+3},c_j)
\big]
\prod_k W_g(s_k,h_k)W_g(s_{k+2N},h_k).
\label{eq:deformed_RBM_II}
\end{align}
\end{widetext}
by introducing $W_g$ which gives
\begin{equation}
\sum_{h=0}^{N-1}
W_g(s,h)W_g(t,h)
=
\Lambda(g)_{st}.
\label{eq:deformation_weight}
\end{equation}
This shows that the RBM construction continues to apply to the cluster states under the deformation.

While the fixed-point states $|\psi^\alpha_0\rangle$ have zero correlation length, the deformations make it finite. We show this by using the normalization condition of the corresponding MPS representations. Let $T^{[\ell_\alpha]}$ be the MPS transfer matrix of the state $|\psi^\alpha_g\rangle$ blocked over the interaction range $\ell_\alpha$. With $M_\alpha=L/\ell_\alpha$, the blocked MPS transfer matrices must satisfy the normalization condition
\begin{align}
{\rm Tr} \left[(T^{[\ell_\alpha]})^{M_\alpha}\right] &= \sum_{\bf s}|\Psi_g^\alpha({\bf s})|^2, \nn
|\Psi_g^\alpha({\bf s})|^2&=\prod_{j=1}^{\ell_\alpha}\prod_{m=1}^{M_\alpha}|\Lambda(g)_{s_{j+(m-1)\ell_\alpha},s_{j+m\ell_\alpha}}|^2.
\label{eq:deform_norm_1}
\end{align}
Define the matrix $W_{s,t}=|\Lambda(g)_{s,t}|^2=e^{g\delta_{s,t}}$, which has the eigenvalues
\begin{equation}
\lambda_0=e^g+N-1,\qquad
\lambda_q=e^g-1\quad(q=1,\ldots,N-1).
\end{equation}
The right-hand side of the normalization condition in Eq.~\eqref{eq:deform_norm_1} can then be written as
\begin{equation}
\sum_{\bf s}|\Psi_g^\alpha({\bf s})|^2 = {\rm Tr}[W^{M_\alpha}]^{\ell_\alpha}= [\lambda_0^{M_\alpha}+(N-1)\lambda_1^{M_\alpha}]^{\ell_\alpha}.
\end{equation}
Applying the binomial formula gives the condition
\begin{equation}
{\rm Tr} \left[(T^{[\ell_\alpha]})^{M_\alpha}\right] = \sum_{k=0}^{\ell_\alpha}\binom{\ell_\alpha}{k} (N-1)^k (\lambda_0^{\ell_\alpha-k}\lambda_1^k)^{M_\alpha} ,
\end{equation}
which must hold for all possible $M_\alpha$. Consequently, the eigenvalues of the blocked MPS transfer matrix are determined as $\mu_k=\lambda_0^{\ell_\alpha-k}\lambda_1^k ~(k=0,\dots,\ell_\alpha)$ up to a phase ambiguity, with multiplicity $m_k=\binom{\ell_\alpha}{k} (N-1)^k$. The unique leading eigenvalue $\mu_0$ and the subleading eigenvalue $\mu_1$ yields the correlation length
\begin{equation}
\xi_\alpha(g)=\frac{\ell_\alpha}{\log[\mu_0/\mu_1]}=\frac{\ell_\alpha}
{\log\!\left[(e^g+N-1)/(e^g-1)\right]}.
\label{eq:deform_correlation_length}
\end{equation}
These examples establish that the $P$-representation can describe exact ground states with nonzero correlation lengths.

\section{Summary}
Neural quantum states have emerged as a powerful framework for encoding many-body quantum ground states, motivated in large part by advances in the AI community. In this work, we revisited the $\mathbb Z_N$ cluster and dipolar cluster states as paradigmatic one-dimensional SPT states protected by spatially uniform and modulated symmetries, and analyzed them with the view of NQS in mind. For all the cluster models analyzed in this work, we derive compact neural-network representations of the wavefunctions both for pristine and deformed cluster Hamiltonians. It is shown that all three cluster states allow the $P$-representation, in which the physical degrees of freedom are encoded solely through a local projector $P^s$, and interaction between spins is mediated by a number of interaction matrices. The NQS representation emerges when we factorize the interaction matrices into a pair of weight matrices. The scheme applies even for deformed cluster states with finite correlation lengths. When each projector carries only two virtual legs, the $P$-representation becomes equivalent to the familiar MPS description of the cluster state. When more than two virtual legs are associated with the projector, the resulting construction is better described as a tensor product state.

\acknowledgments 
JHH was supported by the National Research Foundation of Korea (NRF) grant funded by the Korea government (MSIT) (Grant No. 2023R1A2C1002644 and No. RS-2024-00410027). He thanks \"{O}mer M. Aksoy, Zhian Jia, Nisarga Paul, Shu-Heng Shao, Xiao-Gang Wen, Yizhi You for informative discussion. H.-Y.L was supported by the Basic Science Research Program through the National Research Foundation of Korea funded by the Ministry of Science and ICT [Grant No. RS-2023-00220471, RS-2025-16064392].

\appendix

\section{Proof of symmetry fractionalization in the cluster state}
\label{app:fractionalization-of-cluster-state}

Suppose we performed the $C_1^\dag$ transformation on the open-chain cluster ground state \eqref{CS_open}, to get
\begin{align}
    C_1^\dag|\Psi\rangle =& \mkern-12mu \sum_{s_2,\dots,s_{L-1}} \mkern-12mu \langle s_1 | A^{s_2} \cdots A^{s_{L}} |s_L\rangle |s_1-1,s_2,s_3-1,\dots\rangle \nn
    =& X_1^\dag \mkern-12mu \sum_{s_2,\dots,s_{L-1}} \mkern-12mu \langle s_1 | A^{s_2}A^{s_3+1}A^{s_4} \cdots A^{s_{L-1}+1}A^{s_{L}} |s_L\rangle |\mathbf{s}\rangle.
\end{align} 
Applying the identities $A^{s_{2j-1}+1} = Z^\dag A^{s_{2j-1}} X^\dag$ and $X^\dag A^{s_{2j}} Z^\dag = A^{s_{2j}}$ successively, we are left with
\begin{align}
    C_1^\dag|\Psi\rangle =& X_1^\dag \mkern-12mu \sum_{s_2,\dots,s_{L-1}} \mkern-12mu \langle s_1 | A^{s_2} Z^\dag A^{s_3}A^{s_4} \cdots A^{s_{L}} Z |s_L\rangle |{\bf s}\rangle \nn
    =& X_1^\dag \omega^{-s_2+s_L} \mkern-12mu \sum_{s_2,\dots,s_{L-1}} \mkern-12mu \langle s_1 | A^{s_2}A^{s_3}A^{s_4} \cdots A^{s_{L}} |s_L\rangle |{\bf s}\rangle \nn
    =& X_1^\dag Z_2^\dag Z_L |\Psi\rangle. 
\end{align}
Similarly, performing $C_2^\dag$ on an open-chain state results in
\begin{align} 
    C_2^\dag|\Psi\rangle =& \mkern-12mu \sum_{s_2,\dots,s_{L-1}} \mkern-12mu \langle s_1 | A^{s_2} \cdots A^{s_{L}} |s_L\rangle |s_1,s_2-1,\dots,s_L-1\rangle \nn
    =& X_L^\dag \mkern-12mu \sum_{s_2,\dots,s_{L-1}} \mkern-12mu \langle s_1 | A^{s_2+1}A^{s_3}A^{s_4+1} \cdots A^{s_{L}} |s_L\rangle |\mathbf{s}\rangle \nn
    =& X_L^\dag \mkern-12mu \sum_{s_2,\dots,s_{L-1}} \mkern-12mu \langle s_1 | Z A^{s_2} A^{s_3} A^{s_4} \cdots A^{s_{L-1}} Z^\dag A^{s_{L}} |s_L\rangle |\mathbf{s}\rangle \nn
    =& X_L^\dag \omega^{s_1-s_{L-1}} \mkern-12mu \sum_{s_2,\dots,s_{L-1}} \mkern-12mu \langle s_1 | A^{s_2}A^{s_3}A^{s_4} \cdots A^{s_{L}} |s_L\rangle |{\bf s}\rangle \nn
    =& Z_1 Z_{L-1}^\dag X_L^\dag |\Psi\rangle.
\end{align}

The proof of symmetry fractionalization proceeds in a similar fashion for the P-representation. Acting on the open-chain ground state \eqref{OBC} with $C_1$ yields
\begin{align}
    C_1^\dag|\Psi\rangle =& X_1^\dag \mkern-12mu \sum_{s_2,\dots,s_{L-1}} \mkern-12mu \langle s_1 | \bm\Omega P^{s_2} \bm\Omega^* ( X P^{s_3} X^\dagger ) \cdots \nn
    & \mkern60mu \cdots ( XP^{s_{L-1}} X^\dagger ) \bm\Omega|s_L \rangle |\mathbf{s}\rangle \nn
    =& X_1^\dag \mkern-12mu \sum_{s_2,\dots,s_{L-1}} \mkern-12mu \langle s_1 | \bm\Omega Z^\dag (Z P^{s_2} Z^\dag) \cdots \nn
    & \mkern60mu \cdots \bm\Omega (Z P^{s_{L-2}} Z^\dag) \bm\Omega^* P^{s_{L-1}} \bm\Omega Z|s_L \rangle |\mathbf{s}\rangle \nn
    =& X_1^\dag \mkern-12mu \sum_{s_2,\dots,s_{L-1}} \mkern-12mu \langle s_1 | \bm\Omega Z^\dag P^{s_2}  \bm\Omega^*P^{s_3}\bm\Omega \cdots P^{s_{L-1}}  \bm\Omega Z|s_L \rangle |\mathbf{s}\rangle \nn
    =& X_1^\dag \omega^{-s_2+s_L} \mkern-12mu \sum_{s_2,\dots,s_{L-1}} \mkern-12mu \langle s_1 | \bm\Omega P^{s_2}  \bm\Omega^*P^{s_3}\bm\Omega \cdots P^{s_{L-1}}  \bm\Omega |s_L \rangle |\mathbf{s}\rangle \nn
    =& X_1^\dag Z_2^\dag Z_L |\Psi\rangle. 
\end{align}
A similar exercise gives
\begin{align}\label{C_2wavefunction}
    C_2^\dag|\Psi\rangle =& X_L^\dag \mkern-12mu \sum_{s_2,\dots,s_{L-1}} \mkern-12mu \langle s_1 | \bm\Omega (X P^{s_2} X^\dagger) \cdots \nn
    & \mkern60mu \cdots (X P^{L-2} X^\dag) \bm\Omega^* P^{s_{L-1}}  \bm\Omega |s_L \rangle |\mathbf{s}\rangle \nn 
    =& X_L^\dag \mkern-12mu \sum_{s_2,\dots,s_{L-1}} \mkern-12mu \langle s_1 | Z \bm\Omega  P^{s_2}  \bm\Omega^* (Z^\dag   P^{s_3}  Z) \bm\Omega \cdots \nn 
    & \mkern60mu \cdots (Z^\dag P^{s_{L-1}} Z) Z^\dag \bm\Omega |s_L \rangle |\mathbf{s}\rangle \nn
    =& X_L^\dag \mkern-12mu \sum_{s_2,\dots,s_{L-1}} \mkern-12mu \langle s_1 | Z \bm\Omega P^{s_2} \bm\Omega^* P^{s_3} \bm\Omega \cdots P^{s_{L-1}} Z^\dag \bm\Omega|s_L\rangle |\mathbf{s}\rangle \nn
    =& X_L^\dag \omega^{s_1-s_{L-1}} \mkern-12mu \sum_{s_2,\dots,s_{L-1}} \mkern-12mu \langle s_1 | \bm\Omega P^{s_2} \bm\Omega^* P^{s_3} \bm\Omega \cdots P^{s_{L-1}} \bm\Omega|s_L\rangle |\mathbf{s}\rangle \nn
    =& Z_1 Z_{L-1}^\dag X_L^\dag |\Psi\rangle.
\end{align}

\section{Derivation of weight matrix $W(s,h)$ for the $\Z_N$ cluster state}
\label{app:derivation-of-W}

In this section we introduce a general scheme to construct the function $W$ satisfying $\sum_h W(x,h)W(y,h) =\omega^{pxy}$. Starting from the ansatz
\begin{align}
    W(x,h) = \frac{1}{\sqrt{\kappa}} \omega^{ah^2 + bx^2 + cxh}\,,
\end{align}
we directly get
\begin{align}
    \sum_h W(x,h) W(y,h) = \frac{1}{\kappa} \omega^{b(x^2 + y^2)}\sum_h \omega^{2ah^2 + c(x+y)h}\,.
\end{align}
For some integer $r$ satisfying the equation
\begin{align}\label{condition_square}
    c(x+y)-4ar =0 \mod N\,,
\end{align}
we can complete the square and write
\begin{align}\label{completed_sqaure}
    \sum_h W(x,h) W(y,h) = \frac{1}{\kappa} \omega^{-2ar^2 +b(x^2 + y^2)}\sum_h \omega^{2a(h+r)^2}\,.
\end{align}
The conditions for the periodicity of the summand in Eq.~\eqref{completed_sqaure} are
\begin{align}\label{condition_periodicity}
    2a&\in \mathbb{Z} \quad \text{for odd }N\,, \nonumber
    \\
    4a&\in \mathbb{Z} \quad \text{for even }N\,.
\end{align}
Employing the constant $a$ that ensures periodicity of the summand, Eq.~\eqref{completed_sqaure} becomes
\begin{align}
    \sum_h W(x,h) W(y,h) = \frac{1}{\kappa} \omega^{-2ar^2 +b(x^2 + y^2)} \sum_h \omega^{2ah^2}\,.
\end{align}
Then it is natural to define the normalization factor $\kappa$ that is supposed to be nonzero as
\begin{align}\label{condition_normal}
    \kappa = \sum_h \omega^{2ah^2} \neq 0 \,.
\end{align}
When we further invoke the equation
\begin{align}\label{condition_cross}
    2ar^2 = b(x+y)^2 \mod N \,,
\end{align}
we finally obtain
\begin{align}
    \sum_h W(x,h) W(y,h) = \omega^{-2bxy}\,.
\end{align}

To make the triplet $(a,b,c)$ constant with respect to the arguments $x$ and $y$, we set
\begin{align}
    r = (x+y)k\,,
\end{align}
for some integer $k$. Plugging this into Eq.~\eqref{condition_square} and Eq.~\eqref{condition_cross}, we have
\begin{align}\label{abck}
    c &= 4ak \mod N\,, \nonumber
    \\
    b &=2a k^2  \mod N \,.
\end{align}
Let us assume $k = \pm 1 $ for a simple solution and the conditions Eq.~\eqref{condition_periodicity} and Eq.~\eqref{condition_normal} are satisfied. Then the triplet
\begin{align}
    (a,b,c) = \left(-\frac{p}{4}, -\frac{p}{2}, \pm p \right)
\end{align}
gives
\begin{align}
    \sum_h W(x,h) W(y,h) = \omega^{pxy}\,,
\end{align}
for $p \in \mathbb{Z}_N$, and $W$ constructs an RBM representation of the SPT states associated with $H^2(\mathbb{Z}_N\times\mathbb{Z}_N,U(1))$.

\section{Proof of symmetry invariance for MPS of the cluster state}
\label{app:second-MPS-invariance-proof} 
Next we use the MPS representation in \eqref{A-in-terms-of-W} to prove the invariance, which turns out to be much more involved than the above proof made in the NQS representation. To this end, we need to know how $\W$ transforms with respect to $X$ and $Z$, which are captured by several identities
\begin{align}\label{WX}
    \mathbf{W}^\dagger X &= \omega^{\frac{1}{2}} Z \mathbf{W}^\dagger Z \,, \nn 
    X^\dagger \mathbf{W} &= \omega^{-\frac{1}{2}}  Z^{-1} \mathbf{W} Z^{-1} \,, \nn
    \mathbf{W}^t X &= \omega^{-\frac{1}{2}} Z^{-1} \mathbf{W}^t Z^{-1}\,, \nn 
    X^\dagger \mathbf{W}^* &= \omega^{\frac{1}{2}} Z \mathbf{W}^* Z . 
\end{align}
With these identities the MPSs transform as
\begin{align}
    A^{s_{2j+1}+1 } & =  Z A^{s_{2j+1}}Z^{-1} , \nn 
    A^{s_{2j} +1 } & =  Z^{-1} A^{s_{2j}}Z . 
\end{align}
From the expression of the MPS~\eqref{A-in-terms-of-W} and the identities~\eqref{WX} we show that applying $C_1^\dag$ to the CS wavefunction changes it to 
\begin{align} \label{c1MPS}
\Psi(\mathbf{s}) &\xrightarrow{C_1^\dag} \mathrm{Tr} \left[ \mathbf{W}^\dagger ( X P^{s_1} X^\dagger ) \mathbf{W} \cdot \mathbf{W}^t P^{s_2} \mathbf{W}^* \cdots \mathbf{W}^t P^{s_L}\mathbf{W}^*\right] \nonumber
\\
&=\mathrm{Tr} \!\left[ (Z \mathbf{W}^\dagger P^{s_1}  \mathbf{W} Z^{-1} ) \cdot \mathbf{W}^t P^{s_2} \mathbf{W}^* \!\cdots\! \mathbf{W}^t P^{s_L}\mathbf{W}^*\right] .
\end{align}
It remains to show how, despite these transformations on the MPS matrix, the overall wavefunction stays invariant. Let us first re-organize the expression in Eq.~\eqref{c1MPS} as 
\begin{align}
\mathrm{Tr} \left[ ( \mathbf{W}^* Z \mathbf{W}^\dagger ) P^{s_1}  ( \mathbf{W} Z^{-1} \mathbf{W}^t ) P^{s_2} ( \mathbf{W}^* Z \mathbf{W}^\dagger ) \cdots \W^t P^{s_L}\right] 
\label{modified-trace} 
\end{align}
where the cyclicity of the trace has been used to move $\W^*$ at the end of the trace to the beginning of it. To make further progress we use following identities:
\begin{align}\label{WZW}
    \mathbf{W}Z^{-1} \mathbf{W}^t &= \omega^{-\frac{1}{2}} X^\dagger \mathbf{W}\mathbf{W}^t X = \omega^{-\frac{1}{2}} X^\dagger \bm\Omega X \nonumber
    \\
    \mathbf{W}^*Z \mathbf{W}^\dagger &= \omega^{\frac{1}{2}} X^\dagger \mathbf{W}^*\mathbf{W}^\dagger X = \omega^{\frac{1}{2}} X^\dagger \bm\Omega^* X
\end{align}
Plugging \eqref{WZW} into \eqref{modified-trace} gives
\begin{align}
& \mathrm{Tr} \left[ (X^\dagger\bm\Omega^* X) P^{s_1} (X^\dagger  \bm\Omega X) P^{s_2} (X^\dagger \bm\Omega^* X) \cdots (X^\dagger\bm\Omega X) P^{s_L} \right] \nn 
& \!\! = \mathrm{Tr}  \left[ (\bm\Omega^* Z^\dagger X) P^{s_1} (  \bm\Omega Z X) P^{s_2} (\bm\Omega^* Z^\dagger X) \cdots (\bm\Omega Z X) P^{s_L} \right] \nn
& \!\! = \mathrm{Tr}  \left[ (\bm\Omega^*  \omega^{-1} X Z^\dagger) P^{s_1} (  \bm\Omega \omega XZ ) P^{s_2} (\bm\Omega^* Z^\dagger X) \cdots (\bm\Omega Z X) P^{s_L} \right] \nn
& \!\! = \mathrm{Tr}  \left[ (\bm\Omega^*   X Z^\dagger) P^{s_1} (  \bm\Omega  XZ ) P^{s_2} (\bm\Omega^* X Z^\dagger) \cdots (\bm\Omega XZ ) P^{s_L} \right] \nn
& \!\! = \mathrm{Tr}  \left[ (Z^\dagger \bm\Omega^* Z^\dagger) P^{s_1} (  Z \bm\Omega  Z ) P^{s_2} (Z^\dagger \bm\Omega^*  Z^\dagger) \cdots (Z \bm\Omega Z ) P^{s_L} \right] \nn
& \!\! = \mathrm{Tr}  \left[  \bm\Omega^* (Z^\dagger P^{s_1}   Z) \bm\Omega  (Z  P^{s_2} Z^\dagger) \bm\Omega^*  Z^\dagger \cdots Z \bm\Omega  (Z  P^{s_L} Z^\dagger) \right] \nn
& \!\! = \mathrm{Tr}  \left[  \bm\Omega^*  P^{s_1}  \bm\Omega   P^{s_2} \bm\Omega^*   \cdots \bm\Omega  P^{s_L} \right] = \Psi(\mathbf{s}) 
\end{align}
verifying the invariance of the MPS wavefunction under $C_1$. A similar exercise establishes the invariance under $C_2$.

The identities of the matrix $\mathbf{W}$ and the $\mathbb{Z}_N$ Pauli matrices can be obtained by algebraic calculation of their matrix elements. The first line of Eq.~\eqref{WX} can be explicitly shown by
\begin{align}
    [\mathbf{W}^\dagger X ]_{\alpha\beta} &\equiv \langle\alpha| \mathbf{W}^\dagger X |\beta\rangle = \langle \alpha| \mathbf{W}^\dagger |\beta +1 \rangle \nonumber
    \\
    &=\frac{1}{\sqrt{\kappa}} \omega^{-(a\alpha^2 + b(\beta +1)^2 + c\alpha(\beta+1))} \nonumber
    \\
    &=\frac{1}{\sqrt{\kappa}}\omega^{-(a\alpha^2 + b\beta^2 + c\alpha\beta)} \omega^{-2b\beta} \omega^{-c\alpha} \omega^{-b} \nonumber
    \\
    &= \langle \alpha |\omega^{-b} Z^{-c} \mathbf{W}^\dagger Z^{-2b}  | \beta \rangle \,.
\end{align}
The second line is computed by
\begin{align}
    [X^\dagger \mathbf{W}]_{\alpha\beta}&\equiv  \langle \alpha | X^\dagger \mathbf{W} | \beta \rangle = \langle \alpha+1 | \mathbf{W}|\beta \rangle \nonumber
    \\
    &=\frac{1}{\sqrt{\kappa}} \omega^{a\beta^2 + b (\alpha+1)^2  + c(\alpha+1)\beta} \nonumber
    \\
    &= \frac{1}{\sqrt{\kappa}} \omega^{a \beta^2  + b\alpha^2 + c\alpha \beta +2b\alpha + b + c\beta} \nonumber
    \\
    &= \langle \alpha |\omega^b Z^{2b} \mathbf{W} Z^c |\beta \rangle 
\end{align}

\begin{widetext}
We now explicitly calculate $\mathbf{W}Z^c\mathbf{W}^t$~\eqref{WZW}. Using the coordinate representations $[Z]_{\alpha\beta} = \delta_{\alpha\beta} \omega^\alpha$ and $[Z^c]_{\alpha\beta} = \delta_{\alpha\beta} \omega^{c\alpha}$, we obtain
\begin{align}
    [\mathbf{W}Z^c \mathbf{W}^t]_{\alpha \eta} &= \sum_{\beta \gamma} [\mathbf{W}]_{\alpha\beta} [Z^c]_{\beta \gamma} [\mathbf{W}^t]_{\gamma\eta} =\frac{1}{\kappa} \omega^{b(\alpha^2 + \eta^2)}\sum_{\beta} \omega^{2a\beta^2 + c\beta(\alpha + \eta + 1)}\nonumber
    \\
    &= \omega^{b(\alpha^2 + \eta^2) -b \left(\alpha+\eta+1\right)^2} = \omega^{-2b(\alpha+1)(\eta+1)}\omega^{b}\nonumber
    \\
    &= \omega^{b}\langle \alpha| X^\dagger \mathbf{W} \mathbf{W}^t X |\eta\rangle\,.
\end{align}
Here we use the relation~\eqref{abck}. Similarly, $\mathbf{W}^*Z^{-c} \mathbf{W}^\dagger$ gives
\begin{align}
    [\mathbf{W}^*Z^{-c} \mathbf{W}^\dagger]_{\alpha \eta} &= \sum_{\beta \gamma} [\mathbf{W}^*]_{\alpha\beta} [Z^{-c}]_{\beta \gamma} [\mathbf{W}^\dagger]_{\gamma\eta} = \frac{1}{\kappa^*}\omega^{-b(\alpha^2 +\eta^2)}\sum_\beta \omega^{-2a\beta^2  -c\beta(\alpha+\eta+1)}  \nonumber
    \\
    &= \omega^{-b(\alpha^2 + \eta^2) +b \left(\alpha+\eta+1\right)^2}= \omega^{2b(\alpha+1)(\eta+1)}\omega^{-b} \nonumber
    \\
    &= \omega^{-b}\langle \alpha| X^{\dagger}\mathbf{W}^* \mathbf{W}^\dagger X|\eta\rangle\,.
\end{align}

\section{Proof of symmetry for dCS$_{\rm II}$}
\label{app:symmetry_dCS2}
We prove the NQS representation of the dCS$_{\rm II}$ wavefunction is invariant under $\{C_1, C_2, D_1, D_2\}$. Invoking the relation~\eqref{Wshift}, the wavefunction under $C_1^\dag$ is written as
\begin{align}\label{C1_dCS}
    \Psi(\mathbf{s}) \xrightarrow[]{C^\dagger_1} \prod_j \sum_{\alpha_j \beta_j \gamma_j} & W(s_{2j-1}+1,\alpha_j)W(s_{2j},\alpha_j) \tilde{W}(s_{2j}, \beta_j)\tilde{W}(s_{2j+1}+1, \beta_j) W(s_{2j},\gamma_j)W(s_{2j+3}+1,\gamma_j) \nonumber
    \\
    =\prod_j \sum_{\alpha_j \beta_j \gamma_j}&  \bigg[ W(s_{2j-1},\alpha_j)W(s_{2j},\alpha_j) \tilde{W}(s_{2j}, \beta_j)\tilde{W}(s_{2j+1}, \beta_j) W(s_{2j},\gamma_j)W(s_{2j+3},\gamma_j) \nonumber
    \\
    & \times  \omega^{2b(s_{2j-1}+s_{2j+3}-2s_{2j+1})}\omega^{c(\alpha_j + \gamma_j)} \omega^{\tilde{c} \beta_j} \bigg] \nonumber
    \\
    =\prod_j \sum_{\alpha_j \beta_j \gamma_j}&  W(s_{2j-1},\alpha_j)W(s_{2j},\alpha_j) \tilde{W}(s_{2j}, \beta_j)\tilde{W}(s_{2j+1}, \beta_j) W(s_{2j},\gamma_j)W(s_{2j+3},\gamma_j)\omega^{c(\alpha_j + \gamma_j)} \omega^{\tilde{c} \beta_j}\,.
\end{align}
The summation over the hidden variable $\alpha_j$ is explicitly given by
\begin{align}
    \sum_{\alpha_j}W(s_{2j-1}, \alpha_j) W(s_{2j},\alpha_j)\omega^{c\alpha_j} & = \omega^{-b} \omega^{-2b(s_{2j-1} s_{2j})} \omega^{-2b(s_{2j +1} + s_{2j})}\,.
\end{align}
Similar to this we can explicitly sum over all hidden variables, and the wavefunction~\eqref{C1_dCS} is written as
\begin{align}
    \prod_j \omega^{-2b(s_{2j-1} s_{2j})} \omega^{-2\tilde{b}(s_{2j} s_{2j+1})}\omega^{-2b(s_{2j} s_{2j+3})}\omega^{-2b(s_{2j +1} + 2s_{2j} +s_{2j+3})}\omega^{-2\tilde{b}(s_{2j} + s_{2j+1})}\,. \label{summedoverabg}
\end{align}
Since the constants $b$ and $\tilde{b}$ satisfy
\begin{align}
    \omega^{-2b xy} = \omega^{xy}\,, \quad \omega^{-2\tilde{b}xy} = \omega^{-2xy}\,,
\end{align}
the Eq.~\eqref{summedoverabg} recovers the original wavefunction
\begin{align}
    &\prod_j \omega^{-2b(s_{2j-1} s_{2j})} \omega^{-2\tilde{b}(s_{2j} s_{2j+1})}\omega^{-2b(s_{2j} s_{2j+3})}\omega^{-2b(s_{2j +1} + 2s_{2j} +s_{2j+3})}\omega^{-2\tilde{b}(s_{2j} + s_{2j+1})} \nonumber
    \\
    &=\prod_j \omega^{s_{2j-1} s_{2j}} \omega^{-2s_{2j} s_{2j+1}}\omega^{s_{2j} s_{2j+3}} = \Psi(\mathbf{s})\,.\, 
\end{align}
Proof for $C_2$ is similar.

Now let us consider the dipole symmetry of the dipolar cluster state. Utilizing the relation~\eqref{Wshift}
we show that the wavefunction is invariant under the operation $D_1^\dagger$
\begin{align}
    \Psi(\mathbf{s}) \xrightarrow[]{D^\dagger_1}&  \prod_j \sum_{\alpha_j \beta_j \gamma_j}  W(s_{2j-1}+j,\alpha_j)W(s_{2j},\alpha_j) \tilde{W}(s_{2j}, \beta_j)\tilde{W}(s_{2j+1}+j+1, \beta_j) W(s_{2j},\gamma_j)W(s_{2j+3}+j+2,\gamma_j) \nonumber
    \\
     =& \prod_j \sum_{\alpha_j \beta_j \gamma_j}  \bigg[ W(s_{2j-1},\alpha_j)W(s_{2j},\alpha_j) \tilde{W}(s_{2j}, \beta_j)\tilde{W}(s_{2j+1}, \beta_j) W(s_{2j},\gamma_j)W(s_{2j+3},\gamma_j) \nonumber
    \\
    & \times \omega^{2bjs_{2j-1} + c j \alpha_j + bj^2} \omega^{2\tilde{b}(j+1)s_{2j+1} + \tilde{c} (j+1) \beta_j + \tilde{b}(j+1)^2}  \omega^{2b(j+2)s_{2j+3} + c (j+2) \gamma_j + b(j+2)^2} \bigg] \nonumber
    \\
    =& \prod_j \sum_{\alpha_j \beta_j \gamma_j} \bigg[ W(s_{2j-1},\alpha_j)W(s_{2j},\alpha_j) \tilde{W}(s_{2j}, \beta_j)\tilde{W}(s_{2j+1}, \beta_j) W(s_{2j},\gamma_j)W(s_{2j+3},\gamma_j)    \omega^{c(j\alpha_j +(j+2) \gamma_j -2 (j+1)\beta_j)} \omega^{2b}\bigg] \nonumber
    \\
    =& \prod_j    \omega^{s_{2j-1} s_{2j}} \omega^{-2s_{2j} s_{2j+1}}\omega^{s_{2j} s_{2j+3}} = \Psi(\mathbf{s})\,.
\end{align}
Again a similar proof applies for symmetry under $D_2$.

We can also show that the $P$-representation~\eqref{dCS2-using-projectors} is invariant under $\{C_1, C_2, D_1, D_2\}$. Let us see how the projector representation transforms under $C_2^\dagger$:
\begin{align}\label{rep_omega}
    \Psi(\mathbf{s}) &\xrightarrow[]{C_2^\dagger}\sum_{\bm{\alpha}\bm{\beta}\bm{\gamma} }\sum_{\bm{\alpha}'\bm{\beta}'\bm{\gamma}'}\prod_j P^{s_{2j}+1}_{\alpha'_j\beta'_j\gamma'_j}\Omega_{\alpha'_j\alpha_j}\tilde{\Omega}_{\beta'_j\beta_j}\Omega_{\gamma'_j\gamma_j}P^{s_{2j+1}}_{\alpha_{j+1}\beta_j\gamma_{j-1}} \nonumber 
    \\
    =&\sum_{\bm{\alpha}\bm{\beta}\bm{\gamma} }\sum_{\bm{\alpha}'\bm{\beta}'\bm{\gamma}'}\prod_j P^{s_{2j}}_{\alpha'_j\beta'_j\gamma'_j}\Omega_{(\alpha'_j+1)\alpha_j}\tilde{\Omega}_{(\beta'_j+1)\beta_j}\Omega_{(\gamma'_j+1)\gamma_j}P^{s_{2j+1}}_{\alpha_{j+1}\beta_j\gamma_{j-1}} \nonumber
    \\
    =&\sum_{\bm{\alpha}\bm{\beta}\bm{\gamma} }\sum_{\bm{\alpha}'\bm{\beta}'\bm{\gamma}'}\prod_j P^{s_{2j}}_{\alpha'_j\beta'_j\gamma'_j}\Omega_{\alpha'_j\alpha_j}\tilde{\Omega}_{\beta'_j\beta_j}\Omega_{\gamma'_j\gamma_j}\omega^{\alpha_j -2\beta_j + \gamma_j}P^{s_{2j+1}}_{\alpha_{j+1}\beta_j\gamma_{j-1}} \nonumber 
    \\
    =&\sum_{\bm{\alpha}\bm{\beta}\bm{\gamma} }\sum_{\bm{\alpha}'\bm{\beta}'\bm{\gamma}'}\prod_j P^{s_{2j}}_{\alpha'_j\beta'_j\gamma'_j}\Omega_{\alpha'_j\alpha_j}\tilde{\Omega}_{\beta'_j\beta_j}\Omega_{\gamma'_j\gamma_j}\omega^{\alpha_{j+1} -2\beta_j + \gamma_{j-1}}P^{s_{2j+1}}_{\alpha_{j+1}\beta_j\gamma_{j-1}} \nonumber 
    \\
    =&\sum_{\bm{\alpha}\bm{\beta}\bm{\gamma} }\sum_{\bm{\alpha}'\bm{\beta}'\bm{\gamma}'}\prod_j P^{s_{2j}}_{\alpha'_j\beta'_j\gamma'_j}\Omega_{\alpha'_j\alpha_j}\tilde{\Omega}_{\beta'_j\beta_j}\Omega_{\gamma'_j\gamma_j}P^{s_{2j+1}}_{\alpha_{j+1}\beta_j\gamma_{j-1}}  = \Psi(\mathbf{s})\,.
\end{align}
In going from the first to the second line, we relabel the summation indices as $\alpha'_j \to \alpha'_j+1$, $\beta'_j \to \beta'_j+1$, and $\gamma'_j \to \gamma'_j+1$. In going from the second to the third line, we use the definition of $\Omega$ to extract the factor $\omega^{\alpha_j-2\beta_j+\gamma_j}$. Finally, from the third to the fourth line, we use the property $\prod_j \omega^{\alpha_j}=\prod_j \omega^{\alpha_{j+1}}$.

Similarly, the transformation of the wavefunction under $D_2^\dag$ is cast as
\begin{align}\label{D2_Omega}
    \Psi(\mathbf{s}) &\xrightarrow[]{D_2^\dagger}\sum_{\bm{\alpha}\bm{\beta}\bm{\gamma} }\sum_{\bm{\alpha}'\bm{\beta}'\bm{\gamma}'}\prod_j P^{s_{2j}+j}_{\alpha'_j\beta'_j\gamma'_j}\Omega_{\alpha'_j\alpha_j}\tilde{\Omega}_{\beta'_j\beta_j}\Omega_{\gamma'_j\gamma_j}P^{s_{2j+1}}_{\alpha_{j+1}\beta_j\gamma_{j-1}} \nonumber 
    \\
    =&\sum_{\bm{\alpha}\bm{\beta}\bm{\gamma} }\sum_{\bm{\alpha}'\bm{\beta}'\bm{\gamma}'}\prod_j P^{s_{2j}}_{\alpha'_j\beta'_j\gamma'_j}\Omega_{(\alpha'_j+j)\alpha_j}\tilde{\Omega}_{(\beta'_j+j)\beta_j}\Omega_{(\gamma'_j+j)\gamma_j}P^{s_{2j+1}}_{\alpha_{j+1}\beta_j\gamma_{j-1}} \nonumber
    \\
    =&\sum_{\bm{\alpha}\bm{\beta}\bm{\gamma} }\sum_{\bm{\alpha}'\bm{\beta}'\bm{\gamma}'}\prod_j P^{s_{2j}}_{\alpha'_j\beta'_j\gamma'_j}\Omega_{\alpha'_j\alpha_j}\tilde{\Omega}_{\beta'_j\beta_j}\Omega_{\gamma'_j\gamma_j}\omega^{j(\alpha_j -2\beta_j + \gamma_j)}P^{s_{2j+1}}_{\alpha_{j+1}\beta_j\gamma_{j-1}} \nonumber 
    \\
    =&\sum_{\bm{\alpha}\bm{\beta}\bm{\gamma} }\sum_{\bm{\alpha}'\bm{\beta}'\bm{\gamma}'}\prod_j P^{s_{2j}}_{\alpha'_j\beta'_j\gamma'_j}\Omega_{\alpha'_j\alpha_j}\tilde{\Omega}_{\beta'_j\beta_j}\Omega_{\gamma'_j\gamma_j} \omega^{(j+1)\alpha_{j+1} -2j\beta_j + (j-1)\gamma_{j-1}} P^{s_{2j+1}}_{\alpha_{j+1}\beta_j\gamma_{j-1}} \nonumber 
    \\
    =&\sum_{\bm{\alpha}\bm{\beta}\bm{\gamma} }\sum_{\bm{\alpha}'\bm{\beta}'\bm{\gamma}'}\prod_j P^{s_{2j}}_{\alpha'_j\beta'_j\gamma'_j}\Omega_{\alpha'_j\alpha_j}\tilde{\Omega}_{\beta'_j\beta_j}\Omega_{\gamma'_j\gamma_j}P^{s_{2j+1}}_{\alpha_{j+1}\beta_j\gamma_{j-1}}  = \Psi(\mathbf{s})\,.
\end{align}
Equation~\eqref{D2_Omega} follows by steps entirely analogous to those used in deriving Eq.~\eqref{rep_omega}.

\section{Kramers-Wannier operation as dipolar Fourier transform and its MPO}
\label{sec:KW}

Recent investigation has identified another symmetry of the cluster model which is, unlike the other global and modulated symmetries, non-invertible~\cite{fendley16,shao-TASI23,shao24,yamazaki24,han25,maeda25,pace26,ebisu25,yao25}. The $\Z_N$ Kramers-Wannier (KW) operator implementing such non-invertible symmetry (NIS) is 

\begin{align}
    K = \sum_{\mathbf{g},\mathbf{g}'} \omega^{\sum_j (g_{j}-g_{j+1})g'_{j}}|\mathbf{g}'\rangle\langle\mathbf{g}|,
\label{KW}
\end{align}
with $\mathbf{g}=\{g_1 , g_2 , \cdots \}$ and $\mathbf{g}' = \{ g'_1 , g'_2 , \cdots \}$ being the collection of $\Z_N$ variables. The non-invertible structures of the cluster state are part of the broader framework of non-invertible symmetries recently studied in quantum field theory and condensed matter systems~\cite{sakura22,shao-TASI23}. 

Readers who are familiar with the work of \cite{fendley16} will realize that the expression in \eqref{KW} is the $\Z_N$ generalization of the $N=2$ KW operator introduced there. In \cite{seiberg24}, the authors provided the MPO version of the KW operator, which we can also generalize to $\Z_N$,

\begin{align}
K = {\rm Tr}[\K^1\otimes\cdots\otimes\K^L] = \sum_\mathbf{g} (\K^1)_{g_1,g_2} \otimes \cdots \otimes (\K^L)_{g_L,g_1}\,.
\label{KW-MPO1}
\end{align}
Here, ${\bf g} = \{g_1 , \cdots, g_L \}$, and the local MPO tensor $\K^j$ contains the elements which are themselves operators: 
\begin{align}
 ( \K^j )_{g_j,g_{j+1}} = \sum_{g'_j} \omega^{(g_j-g_{j+1})g'_j} |g'_j\rangle\langle g_j| = |\overline{g_{j+1}-g_j} \rangle \langle g_j | . 
\label{KW-MPO2}
\end{align}
The overlined state $|\overline{g} \rangle =  \sum_s \omega^{-sg} |s\rangle$ is the eigenstate of local $X$ operator $X|\overline{g}\rangle = \omega^g |\overline{g} \rangle$. One can check that the expression in \eqref{KW-MPO2} reduces to the one in \cite{seiberg24} by taking $N=2$. The MPO construction of the KW transformation is closely related to the general gauging framework of Haegeman \textit{et al.}~\cite{Haegeman2015}

Viewed as the action on a given wavefunction $\Psi({\bf s})$, the KW operator implements the transformation
\begin{align} \Psi(\mathbf{s}) \xrightarrow{K} \Psi' ({\bf s} ) = \sum_{\mathbf{g}}\omega^{\sum_j g_{j} (s_{j} - s_{j-1} ) }\Psi(\mathbf{g}) .
\label{ZN-KW} 
\end{align} 
This is reminiscent of the Fourier transformation of the function $\Psi ({\bf g})$, except for the subtle difference that the target variable is the difference $s_{j}-s_{j-1}$ rather than $s_j$ itself. In fact, this observation leads to very natural interpretation of the KW transformation as the generalization of the ordinary Fourier transformation in which the target variable is not the {\it charge} of the target space $s_j$, but the {\it dipole} in the same space given by $d_j \equiv s_j - s_{j-1}$. In a sense, the KW transformation is like the {\it dipole Fourier transform} while the original Fourier transform would now be the {\it charge Fourier transform}. The dipolar Fourier transform is non-invertible because the target variable $d_j \equiv s_j - s_{j-1}$ has a constraint, $\sum_j d_j = 0$, and covers only $1/N$-th of the target space. The transformed function $\Psi' ({\bf s})$ must automatically satisfy $\Psi' (\{ s_j \rightarrow s_j + 1 \} ) = \Psi' ({\bf s})$, i.e. become an eigenstate of $C=\prod_j X_j$ with eigenvalue +1. 

Further generalization of this observation is to the {\it quadrupole Fourier transformation} in which the target variable is $q_j = s_{j+1}-2s_j + s_{j-1}$:
\begin{align} \Psi(\mathbf{s}) \rightarrow \Psi' ({\bf s} ) = \sum_{\mathbf{g}}\omega^{\sum_j g_{j} q_j }\Psi(\mathbf{g}) .
\end{align} 
Now the target space has two constraints, in the form of the total ``charge" and the total ``dipole" conservation: $\sum_j q_j = \sum_j j q_j = 0$~\footnote{More precisely, the total dipole moment must be zero modulus $N$, assuming that the length of the chain $L$ is divisible by $N$.}. The transformed function is invariant under both $s_j \rightarrow s_j +1$ and $s_j \rightarrow s_j + j$, i.e. resides in the space where both $C$ and $D=\prod_j (X_j)^j$ are effectively one. The quadrupole Fourier transformation appears as the KW operator implementing the NIS of the dCS$_{\rm II}$ - see App.~\ref{app:NIS-dCSII}.

The NIS operator specific to the $\Z_N$ cluster state is
\begin{align} K_c =TK_{1} K_{2}, 
\end{align} 
where $T=\sum_{\mathbf{g}} \bigotimes_j |g_j\rangle_{j+1}\langle g_j|_j$ gives the translation by one site, and $K_{1}$ and $K_{2}$ are the KW operators acting on the odd and even sublattices, 
\begin{align}
K_{1} & = \sum_{\mathbf{g}_1,\mathbf{g}'_1} \omega^{\sum_n (g_{2n-1}-g_{2n+1})g'_{2n-1}}|\mathbf{g}'_1 \rangle\langle\mathbf{g}_1 |, \nn
K_{2} & = \sum_{\mathbf{g}_2 ,\mathbf{g}'_2} \omega^{\sum_n (g_{2n}-g_{2n+2})g'_{2n}}|\mathbf{g}'_2 \rangle\langle\mathbf{g}_2 |,
\end{align}
with ${\bf g}_1 = \{g_1 , g_3 , \cdots g_{L-1} \}$ and ${\bf g}_2 = \{g_2 , g_4 , \cdots g_L \}$, respectively, for even $L$. Explicitly, 
\begin{align}
K_c = \sum_{\mathbf{g},\mathbf{g}'} \omega^{\sum_j (g_{j-1}-g_{j+1})g'_j}|\mathbf{g}'\rangle\langle\mathbf{g}| ,
\end{align}
where ${\bf g}, {\bf g}'$ span all the qudits of the lattice. Invariance of the CS wavefunction under $K_c$ is easily checked: 
\begin{align} \Psi(\mathbf{s}) \xrightarrow{K_c} \sum_{\mathbf{g}}\omega^{\sum_j  (s_{j+1} - s_{j-1} ) g_{j} }\Psi(\mathbf{g}) \propto \Psi (\mathbf{s}) .
\end{align} 
One can see that this is dipole Fourier transformation to $d_j \equiv s_{j+1}-s_{j-1}$. The new variable satisfies $\sum_{j \in {\rm odd}} d_j = \sum_{j \in {\rm even}} d_j = 0$ over the odd and even sublattices, separately. Functions that emerge from the $K_c$ mapping automatically satisfy $C_1 = C_2 = 1$. For completeness, we list the fusion rules satisfied by the three symmetry generators $\{K_c, C_1 , C_2 \}$ of the cluster state~\cite{han25}: 
\begin{gather}
    C_1K_c=K_cC_1=K_c, \quad C_2K_c=K_cC_2=K_c, \nn
    K_c^\dag K_c = K_c^2 = \left( \sum_{m=1}^N C_1^m \right) \left(\sum_{n=1}^N C_2^n  \right).
\label{CS-fusion}
\end{gather}

The MPO of the KW operator $K_c$ finds a compact expression in the $P$-representation. We first note that the elements of $K_c$ are
\begin{align}
\label{K_c-elements} 
\langle {\bf s}' | K_c |{\bf s}\rangle & = \omega^{\sum_j (s_{j-1}-s_{j+1})s'_j}  \nn 
& = \omega^{(s_1 - s_3 )s'_2 + (s_3 - s_5 ) s'_4 + \cdots + (s_{L-1} - s_1 ) s'_L } \nn 
& \times \omega^{(s_L - s_2) s'_1 + (s_2 - s_4 ) s'_3 + \cdots + (s_{L-2} -s_L ) s'_{L-1} } . 
\end{align} 
The expression in the first line of \eqref{K_c-elements} has been deliberately split into two parts over the next two lines, each line corresponding to a connected curve going $s_1 \rightarrow s'_2 \rightarrow s_3 \rightarrow \cdots$ or $s'_1 \rightarrow s_2 \rightarrow s'_3 \rightarrow \cdots$ in Fig.~\ref{fig:KW-as-TN}. Each curve can be expressed as a trace:
\begin{align}
\omega^{(s_1 - s_3 )s'_2 + (s_3 - s_5 ) s'_4 + \cdots + (s_{L-1} - s_1 ) s'_L } & =  {\rm Tr}[ {\bm \Omega}^* P^{s_1} {\bm \Omega} P^{s'_2} {\bm \Omega}^* P^{s_3} \cdots {\bm \Omega} P^{s'_L}  ] \nn 
\omega^{(s_L - s_2) s'_1 + (s_2 - s_4) s'_3 + \cdots + (s_{L-2} -s_L ) s'_{L-1} } & = {\rm Tr}[ {\bm \Omega} P^{s'_1} {\bm \Omega}^* P^{s_2} {\bm \Omega} P^{s'_3} \cdots {\bm \Omega}^* P^{s_L} ] .
\end{align}
Combining the two expressions gives
\begin{align} 
\langle {\bf s}' | K_c |{\bf s}\rangle & = {\rm Tr}[ {\bm \Omega}^* P^{s_1} {\bm \Omega} P^{s'_2} {\bm \Omega}^* P^{s_3} \cdots {\bm \Omega} P^{s'_L} ] \times  {\rm Tr}[ {\bm \Omega} P^{s'_1} {\bm \Omega}^* P^{s_2} {\bm \Omega} P^{s'_3} \cdots {\bm \Omega}^* P^{s_L} ] . 
\label{K_c-as-two-traces} 
\end{align}  
Each trace is in fact a CS wavefunction or its complex conjugate, written in a sort of {\it doubled} Hilbert space of $\bf s$ and $\bf s'$. Based on this observation and the fact that each CS wavefunction is invariant under the $C_1$ and $C_2$ symmetries, it follows that $K_c$ is also invariant under each of them. This explains the first two fusion rules in \eqref{CS-fusion}. Invoking the theorem 
\begin{align*} {\rm Tr}[A_1 A_2 \cdots A_L] {\rm Tr}[B_1 B_2 \cdots B_L]= {\rm Tr}[ (A_1 \otimes B_1 ) (A_2 \otimes B_2) \cdots (A_L \otimes B_L) ]
\end{align*} 
allows us to reduce the two-trace expression in Eq.~\eqref{K_c-as-two-traces} to a single trace:
\begin{align}
\langle {\bf s}' | K_c |{\bf s}\rangle & = {\rm Tr}[ ( {\bm \Omega}^* P^{s_1} \otimes {\bm \Omega} P^{s'_1} ) (  {\bm \Omega} P^{s'_2} \otimes {\bm \Omega}^* P^{s_2} ) ( {\bm \Omega}^* P^{s_3} \otimes {\bm \Omega} P^{s'_3} ) \cdots ( {\bm \Omega} P^{s'_L} \otimes {\bm \Omega}^* P^{s_L} ) ] . 
\label{K_c-as-one-trace}     
\end{align}
Since the MPO now involves a pair of projectors at each site, we refer to it as the $P^2$-representation of the MPO for the KW operator. Figure \ref{fig:KW-as-TN} illustrates the MPO of the KW operator $K_c$ graphically. 
\begin{figure}[t]
    \centering
    \includegraphics[width=0.3\columnwidth]{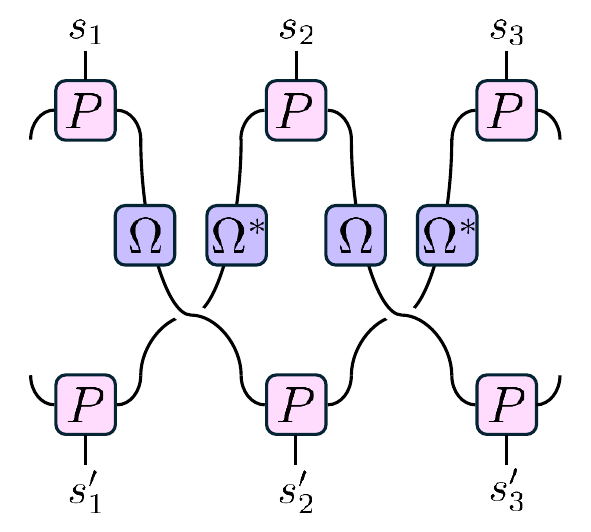}
    \caption{$P^2$-representation of the $K_c$ operator as an MPO.}
    \label{fig:KW-as-TN}
\end{figure}

We can write the KW operator itself, rather than its matrix elements, as an MPO:
\begin{align}
K_c = {\rm Tr}[ \K^1_c \otimes \cdots \otimes \K^L_c ] =\sum_\mathbf{\bm\alpha} (\K^1_c)_{\alpha_L,\alpha_1} \otimes \cdots \otimes (\K^L_c)_{\alpha_{L-1},\alpha_L} , 
\end{align}
where ${\bm \alpha}= \{\alpha_1, \cdots, \alpha_L \}$ and each bond index $\alpha_j  \equiv (g_j , h_j )$ consists of a pair of $\Z_N$ numbers. Each element of the matrix $\K^j_c$ is an operator acting on the qudit at the site $j$. To match the decomposition \eqref{K_c-as-one-trace} of $K_c$, we invoke the identification
\begin{align}
    \langle s'_j|(\K^{j}_c)_{\alpha_{j-1} ,\alpha_j }|s_j\rangle =& ({\bm \Omega}^* P^{s_j} \otimes {\bm \Omega} P^{s'_j})_{\alpha_{j-1} ,\alpha_j} \nn
    =& ({\bm \Omega}^* P^{s_j})_{g_{j-1},g_j} ({\bm \Omega} P^{s'_j})_{h_{j-1},h_j} \nn
    =& \omega^{-g_{j-1}g_j}\delta_{g_j,s_j}\omega^{h_{j-1}h_j}\delta_{h_j,s'_j}
\end{align}
for odd $j$ and
\begin{align}
    \langle s'_j|(\K^{j}_c)_{\alpha_{j-1} ,\alpha_j }|s_j\rangle =& ({\bm \Omega} P^{s'_j} \otimes {\bm \Omega}^* P^{s_j})_{\alpha_{j-1} ,\alpha_j} \nn
    =& ({\bm \Omega} P^{s'_j})_{g_{j-1},g_j} ({\bm \Omega}^* P^{s_j})_{h_{j-1},h_j} \nn
    =& \omega^{g_{j-1}g_j}\delta_{g_j,s'_j}\omega^{-h_{j-1}h_j}\delta_{h_j,s_j}
\end{align}
for even $j$. This leads to the definition
\begin{align}
    (\K^{j}_c)_{\alpha_{j-1} ,\alpha_j } = 
    \begin{cases}
        \omega^{h_{j-1}h_j-g_{j-1}g_j}|h_j\rangle\langle g_j|, & j~{\rm odd}, \\
        \omega^{g_{j-1}g_j-h_{j-1}h_j}|g_j\rangle\langle h_j|, & j~{\rm even}.
    \end{cases}
\end{align}

It turns out that the MPO representation of the KW operator comes in many different forms. Introducing the $\Z_N$ swap operator
\begin{align}
    S = \frac{1}{N} \sum_{a, b=1}^N  (X^a Z^b)\otimes (X^a Z^b)^\dagger,
\label{swap}
\end{align}
we can rewrite Eq. \eqref{K_c-as-one-trace} as
\begin{align}
    \langle {\bf s}' | K_c |{\bf s}\rangle & = {\rm Tr}[ ( {\bm \Omega}^* P^{s_1} \otimes {\bm \Omega} P^{s'_1} ) S (  {\bm \Omega}^* P^{s_2} \otimes {\bm \Omega} P^{s'_2} ) S \cdots ( {\bm \Omega}^* P^{s_L} \otimes {\bm \Omega} P^{s'_L} ) S ]
\end{align}
which is represented by the site-independent local MPO tensor
\begin{align}
    (\K^j_c)_{\alpha_{j-1},\alpha_j} &= \sum_{s_j,s'_j} [( {\bm \Omega}^* P^{s_j} \otimes {\bm \Omega} P^{s'_j} ) S]_{\alpha_{j-1},\alpha_j} |s'_j\rangle\langle s_j| \nn
    &= \omega^{h_{j-1}g_j-g_{j-1}h_j}|g_j\rangle\langle h_j|.
\end{align}

Recalling the NQS decomposition ${\bm \Omega} = \W \W^t, {\bm \Omega}^* = \W^* \W^\dag$, one can re-organize Eq.~\eqref{K_c-as-two-traces} as
\begin{align} 
\langle {\bf s}' | K_c |{\bf s}\rangle & = {\rm Tr}[ (\W^\dag P^{s_1} \W ) ( \W^t P^{s'_2} \W^* ) ( \W^\dag P^{s_3} \W ) \cdots ( \W^t P^{s'_L} \W^* ) ] \times {\rm Tr}[  (\W^t P^{s'_1} \W^* ) ( \W^\dag P^{s_2} \W ) ( \W^t P^{s'_3} \W^* ) \cdots ( \W^\dag P^{s_L} \W ) ] \nn 
& = {\rm Tr}[ ( \W^\dag P^{s_1} \W \otimes \W^t P^{s'_1} \W^* )  ( \W^t P^{s'_2} \W^* \otimes \W^\dag P^{s_2} \W ) \cdots ( \W^t P^{s'_L} \W^* \otimes \W^\dag P^{s_L} \W ) ] \nn 
& = {\rm Tr}[ (\W^\dag \otimes \W^t )(P^{s_1} \otimes P^{s'_1} ) (\W \otimes \W^*) \cdots (\W^t \otimes \W^\dag )(P^{s'_L} \otimes P^{s_L}) ( \W^* \otimes \W ) ] \nn
& = {\rm Tr}[ (\W^\dag \otimes \W^t )(P^{s_1} \otimes P^{s'_1} ) (\W \otimes \W^*) S \cdots S (\W^\dag \otimes \W^t )(P^{s_L} \otimes P^{s'_L} ) (\W \otimes \W^*) S ].
\end{align}  
This gives rise to another MPO representation of $K_c$ with the local tensor
\begin{align}
    (\K^j_c)_{\alpha_{j-1},\alpha_j} &= \sum_{s_j,s'_j}[(\W^\dag \otimes \W^t )(P^{s_j} \otimes P^{s'_j} ) (\W \otimes \W^*) S]_{\alpha_{j-1},\alpha_j} |s'_j\rangle\langle s_j| \nn
    &= \sum_{s_j,s'_j} W^*(s_j,g_{j-1}) W(s'_j,h_{j-1}) W(s_j,h_j) W^*(s'_j,g_j) |s'_j\rangle\langle s_j|.
\end{align}
\end{widetext}

A substantially different approach to the MPO representation of $K_c$ can be made by starting from the MPO representation of the $\Z_N$ KW operator in Eq. \eqref{KW-MPO2}. Invoking the definition of the KW operator for the cluster state, $K_c=TK_1 K_2$, we get the MPO for $K_c$ as
\begin{align}
    K_c = T \sum_{\bf g} \bigotimes_j (\K^j)_{g_j,g_{j+2}} = \sum_{\bf g} \bigotimes_j (\K^j_c)_{g_{j-1},g_{j+1}}, 
\end{align}
where
\begin{align}
    (\K^j_c)_{g_{j-1},g_{j+1}} &= T(\K^{j-1})_{g_{j-1},g_{j+1}} \nn
    &=|\overline{g_{j+1}-g_{j-1}} \rangle_j \langle g_{j-1} |_{j-1}.
\label{K_c-MPO-from-KW1}
\end{align}
Note that each $\K^j_c$ contains $N\times N$ elements, but each element is an operator defined over the two adjacent qudits at $j$ and $j-1$. As a result, we have $K_c$ written as a product of two traces:
\begin{align}
K_c & = {\rm Tr} [\K^1_c \otimes \K^3_c \otimes \cdots \otimes \K^{L-1}_c] \nn 
& \times {\rm Tr} [\K^2_c \otimes \K^4_c \otimes \cdots \otimes \K^L_c] .
\end{align}
This can be re-written as a single trace $$K_c = {\rm Tr} [\K^{(1,2)}_c \otimes \K^{(3,4)}_c \cdots \otimes \K^{(L-1,L)}_c], $$ where the two adjacent sites are now grouped as one effective site and 
\begin{align}
 & (\K^{(2n-1,2n)}_c)_{(g_{2n-2},g_{2n-1}),(g_{2n},g_{2n+1})} \nn 
 & = (\K^{2n-1}_c)_{g_{2n-2},g_{2n}} \otimes (\K^{2n}_c)_{g_{2n-1},g_{2n+1}}.
\end{align}

Finally, we present yet another way to arrive at the MPO representation of $K_c$ by examining how the MPS representation of the cluster state wavefunction transforms under it:
\begin{align}
   \Psi({\bf s}) \xrightarrow{K_c} {\rm Tr} \left[ \prod_{j=1}^L  \left( \sum_{g_j } \omega^{g_j (s_{j+1}-s_{j-1})} A^{g_j} \right)  \right] 
\end{align}
Performing the sum over each $g_j$ gives
\begin{align}
   \sum_{g_j \in \Z_N } \omega^{g_j (s_{j+1}-s_{j-1})} A^{g_j} = \sum_{s'_j} |s'_j \rangle \langle \overline{ s_{j+1}-s_{j-1} + (-1)^j s'_j } |
\label{CS-MPS-under-K_c}
\end{align}
Note that the phase factor $\omega^{g_j (s_{j+1}-s_{j-1})}$ in this transformation can be understood as
\begin{align}
    \omega^{g_j (s_{j+1}-s_{j-1})} =  \langle s_{j-1}| (\K^j_c)_{s_{j-1},s_{j+1}}|g_j\rangle
\end{align}
where the operator $(\K^j_c)_{s_{j-1},s_{j+1}}$ is defined as
\begin{align}
(\K^j_c)_{s_{j-1},s_{j+1}} & = \sum_{g_j} \omega^{(s_{j+1}-s_{j-1})g_j} |s_{j-1} \rangle_{j-1} \langle g_j|_j \nn
& = | s_{j-1} \rangle_{j-1} \langle \overline{s_{j+1}-s_{j-1}} |_j. 
\end{align}
It turns out that this gives rise to another MPO representation of $K_c$ that resembles Eq. \eqref{K_c-MPO-from-KW1}:
\begin{align}
    K_c = \sum_{\bf s} \bigotimes_j (\K^j_c)_{s_{j-1},s_{j+1}}.
\end{align}

Although the discussion of KW symmetry lies somewhat outside the main focus of the paper, we include it here for three reasons. First, it extends the previously known $\Z_2$ MPO representation of the KW operator~\cite{seiberg24} to $\Z_N$. Second, it illustrates the utility of the $P$-representation not only for expressing cluster states themselves, but also for constructing NIS operators as MPO. Finally, we arrive at an intuitively appealing interpretation of the KW transformation as the dipolar version of the discrete Fourier transform with the obvious restriction on the target space giving clear intuition to the non-invertibility of the mapping.

\section{Non-invertible symmetry of dCS$_{\rm II}$}
\label{app:NIS-dCSII}

The type-II dipolar cluster model admits the NIS~\cite{han25}
\begin{align}
    K_d =& \Big(\prod_j S_{2j,2j+1}\Big) \tilde K_1 \tilde K_2^\dag \nn
    =& \sum_{\mathbf{s},\mathbf{s}'} \Phi(\mathbf{s},\mathbf{s}')\Phi^*(\mathbf{s}',\mathbf{s})|\mathbf{s}'\rangle\langle\mathbf{s}|, \nn
    \Phi(\mathbf{s},\mathbf{s}') =& \omega^{\sum_j s_{2j+1}(s'_{2j+2}-2s'_{2j}+s'_{2j-2})} \nn
    =& \omega^{\sum_j (s_{2j-1}-2s_{2j+1}+s_{2j+3})s'_{2j}} . 
\end{align}
Here, $\tilde K_1$ and $\tilde K_2$ are the dipolar KW operators or the quadrupole Fourier transformations 
\begin{align}
    \tilde K = \sum_{\mathbf{s},\mathbf{s}'} \omega^{\sum_j (s_{j+1}-2s_j+s_{j-1})s'_j}|\mathbf{s}'\rangle\langle\mathbf{s}|
\end{align}
acting on the odd and even sublattices, respectively. The swap operator $S_{2j,2j+1}$, already introduced in \eqref{swap}, interchanges the qudit states at $2j$ and $2j+1$.

The $K_d$-invariance of the dCS$_{\rm II}$ ground state can be shown as follows:
\begin{align}
    \Psi(\mathbf{s})\xrightarrow{K_d}& \sum_{\bf g} \Phi(\mathbf{g},\mathbf{s})\Phi^*(\mathbf{s},\mathbf{g}) \Psi(\mathbf{g}) \nn
    &= \sum_{\mathbf{g}_1} \omega^{\sum_j g_{2j+1} (s_{2j+2} -2 s_{2j} + s_{2j-2})} \nn
    & \mkern21mu \times \sum_{\mathbf{g}_2} \omega^{\sum_j g_{2j}(g_{2j-1} - 2g_{2j+1} + g_{2j+3}-s_{2j-1} + 2s_{2j+1} - s_{2j+3})} \nn
    &\propto \sum_{k,\ell\in\Z_N} \omega^{\sum_j (s_{2j+1}+k+\ell j) (s_{2j+2} -2 s_{2j} + s_{2j-2})} \nn
    &\propto \Psi(\mathbf{s}).
\end{align}
In the third line, we used that the summation over the even sites $\mathbf{g}_2=(g_2,g_4,\dots,g_L)$ gives the constraint
\begin{align*}
    &g_{2j-1} - 2g_{2j+1} + g_{2j+3} \equiv s_{2j-1} - 2s_{2j+1} + s_{2j+3} ~({\rm mod} ~N) \\
    & \Longleftrightarrow g_{2j-1}=s_{2j-1}+k+\ell j,~~ k,\ell\in\Z_N.
\end{align*}
The definition of $\Phi({\bf s},{\bf s}')$ makes clear that $K_d$ is a quadrupole Fourier transformation whose image lies entirely in the sector $C_1=C_2=D_1=D_2=1$.

Since $\Phi({\bf s},{\bf s}')=\Psi(s_1,s'_2,s_3,\dots,s'_L)$, each matrix element of $K_d$ is nothing but the product of the dCS$_{\rm II}$ wavefunction and its complex conjugate,
\begin{align}
    \langle {\bf s}'|K_d|{\bf s}\rangle = \Psi(s_1,s'_2,s_3,\dots,s'_L)\Psi^*(s'_1,s_2,s'_3,\dots,s_L).
\end{align}
This shows that the $K_d$-invariance of the dCS$_{\rm II}$ wavefunction immediately implies the first four of the following fusion rules satisfied by the symmetry generators $\{K_d,C_1,C_2,D_1,D_2\}$~\cite{han25}:
\begin{gather}
    C_1K_d=K_dC_1=K_d, \quad C_2K_d=K_dC_2=K_d, \nn
    D_1K_d=K_dD_1=K_d, \quad D_2K_d=K_dD_2=K_d, \nn
    K_d^\dag K_d = K_d^2 = \left( \sum_{m=1}^N C_1^m \right) \left(\sum_{n=1}^N C_2^n  \right)\left( \sum_{p=1}^N D_1^p \right) \left(\sum_{q=1}^N D_2^q  \right).
\end{gather}

Using the TPS representations of the dCS$_{\rm II}$ wavefunction \eqref{dCSII-TPS-P} and \eqref{dCSII-TPS-W}, it is possible to write an MPO representation of $K_d$,
\begin{align}
    K_d = {\rm Tr} [\K_d^{(L,1)} \otimes \K_d^{(2,3)} \otimes \cdots \otimes \K_d^{(L-2,L-1)}],
\end{align}
where the local tensor over each pair of adjacent sites $(2n,2n+1)$ is defined as
\begin{widetext}
\begin{align}
    \big(\K_d^{(2n,2n+1)}\big)_{\Gamma_{n-1},\Gamma_n} = \sum_{\substack{s_{2n},s_{2n+1} \\ s'_{2n},s'_{2n+1}}} \sum_{\beta,\beta'} 
    T^{s'_{2n}}_{\alpha_{n-1}\beta\gamma_{n}} 
    T^{s_{2n+1}}_{\alpha_{n}\beta\gamma_{n-1}} 
    \Big(T^{s_{2n}}_{\alpha'_{n-1}\beta'\gamma'_{n}}\Big)^* 
    \Big(T^{s'_{2n+1}}_{\alpha'_{n}\beta'\gamma'_{n-1}}\Big)^* |s'_{2n},s'_{2n+1}\rangle\langle s_{2n},s_{2n+1}|,
\end{align}
\end{widetext}
and each bond index $\Gamma_j=(\alpha_j,\gamma_j,\alpha'_j,\gamma'_j)$ is a collection of four $\Z_N$ numbers.

\bibliography{CS-NQS}

\end{document}